\documentclass[
 groupedaddress,
 amsmath,
 amssymb,
 aps,
 pra,
 floatfix,
 twocolumn,
 longbibliography
]{revtex4-2}

\newcommand{\bra}[1]{\langle #1\rvert}
\newcommand{\ket}[1]{\lvert #1\rangle}
\newcommand{\ip}[2]{\bra{#1} #2\rangle}
\newcommand{\op}[2]{\ket{#1} \bra{#2}}
\newcommand{\exv}[1]{\langle #1\rangle}

\newcommand{\pd}[1]{\frac{\partial #1}{\partial t}}

\newcommand{\SU}[1]{\mathrm{SU{(#1)}}}
\newcommand{\su}[1]{\mathfrak{su}{(#1)}}
\newcommand{\mh}{\mathcal{H}}
\newcommand{\up}{\ket{\uparrow}}
\newcommand{\upp}{\uparrow}
\newcommand{\down}{\ket{\downarrow}}
\newcommand{\ddown}{\downarrow}

\DeclareMathOperator{\Tr}{Tr}

\usepackage[varg]{newtxmath}
\usepackage{newtxtext}

\usepackage{graphicx}
\usepackage{dcolumn}
\usepackage{bm}
\usepackage{hyperref}

\usepackage{tikz}
\usetikzlibrary{calc}

\usepackage{orcidlink}

\usepackage{amsmath}
\usepackage{upgreek}
\usepackage{physics}
\usepackage{placeins}
\usepackage{comment}
\usepackage{mathrsfs}

\usepackage{ytableau}
\ytableausetup{centertableaux,smalltableaux}
\usepackage{mathdots}
\usepackage{enumitem}

\usepackage[plainruled,vlined]{algorithm2e}
\usepackage{ragged2e}
\makeatletter

\SetAlgoCaptionLayout{RevTeXAlgoCaptionLayout}
\makeatother

\usepackage[capitalise]{cleveref} 
\crefformat{equation}{Eq.~(#2#1#3)} 
\crefformat{section}{Sec.~#2#1#3} 
\Crefformat{equation}{Equation~(#2#1#3)}
\crefformat{figure}{Fig.~#2#1#3}
\crefrangeformat{equation}{Eqs.~#3(#1)#4--#5(#2)#6}
\Crefformat{section}{Section~#2#1#3}

\begin{document}

\title{Efficient Polynomial-Scaled Determination of Algebraic Entanglement Entropy Between Collective Degrees of Freedom}
\author{John Drew Wilson\orcidlink{0000-0001-6334-2460}}
\thanks{These authors contributed equally to this work. Corresponding author: Jarrod.Reilly@colorado.edu}
\affiliation{JILA and Department of Physics, University of Colorado, 440 UCB, Boulder, CO 80309, USA}
\author{Jarrod T. Reilly\orcidlink{0000-0001-5410-089X}}
\thanks{These authors contributed equally to this work. Corresponding author: Jarrod.Reilly@colorado.edu}
\affiliation{JILA and Department of Physics, University of Colorado, 440 UCB, Boulder, CO 80309, USA}
\author{Murray J. Holland\orcidlink{0000-0002-3778-1352}}
\affiliation{JILA and Department of Physics, University of Colorado, 440 UCB, Boulder, CO 80309, USA}



\begin{abstract}
In this work, we explore physical systems which support not only multipartite \emph{inter}particle entanglement, but also \emph{intra}particle entanglement between different degrees of freedom of the constituent particles and entanglement between different degrees of freedom of different particles, i.e., \emph{algebraic} entanglement.
We derive a simple method for calculating the algebraic entanglement entropy between two of the particles' degrees of freedom from collective states of the whole ensemble.
Our procedure makes use of underlying symmetries in these systems, in particular permutation symmetry of the particle indices, and shows a connection between the algebraic entanglement entropy in these systems and the irreducible representations of Lie groups which describe the particles' degrees of freedom.
Namely, we use the direct sum over irreducible representations to diagonalize the reduced density matrices in a block-by-block manner, then utilize the multiplicity of these irreducible representations to reproduce the results from an exponentially-scaled Hilbert space in only polynomial complexity. 
We use this to explore a variety of systems where the constituent particles support two degrees of freedom each with two levels, such as atoms with two electronic states and two momentum states.
Notably, these systems may be exactly simulated in a polynomial-scaled Hilbert space, yet they support an algebraic entanglement entropy that grows linearly with the particle number which typically requires an exponentially-scaled Hilbert space.
\end{abstract}

{
\let\clearpage\relax
\maketitle
}

\section{Introduction} \label{sec:Intro}
Quantum entanglement is one of the most striking features of quantum mechanics with no classical analog.
This makes technologies that utilize quantum entanglement promising for use in improved sensing~\cite{Christopher_Wilsons_PaperAnd_Definitely_Not_Jarrods_Or_Johns,Degan,Tse,PedrozoPenafiel,Robinson,Gilmore,Reilly3}, computing~\cite{Kendon,Bennett,Schaetz,Yin,Liu}, simulation~\cite{GarciaPerez,Kokail,Sundar,Joshi,Euler}, and more~\cite{Horodecki,Gauger,Genoni,Ono,Brunner,Apte,Ecker,Shankar,Liu2}.
Entanglement is crucial to these technologies because non-local correlations between subsystems can be exploited to surpass the capabilities of classical devices, such as overcoming the standard quantum limit~\cite{Christopher_Wilsons_PaperAnd_Definitely_Not_Jarrods_Or_Johns,Tse,PedrozoPenafiel} or teleporting information from a source to destination~\cite{Bennett2,Li,Chen}. 
Due to its importance, many theoretical techniques have been developed to detect, characterize, and analyze the entanglement in a quantum system, e.g., quantum Fisher information~\cite{Christopher_Wilsons_PaperAnd_Definitely_Not_Jarrods_Or_Johns,Hyllus,Liu3}, quantum state tomography~\cite{Smithey,DAuria,Bae}, and the Schmidt decomposition~\cite{Horodecki,Ekert,NielsenChaung}.

Of primary interest in this work is the concept of entanglement entropy~\cite{Horodecki,NielsenChaung,Eisert} which characterizes the degree of entanglement between two subsystems of a quantum device and is of fundamental interest to quantum theory~\cite{Calabrese,Eisert,Kitaev}.
The most simple quantum system which supports entanglement is a bipartite system, where the whole may be divided into two composite systems.
Here, calculating the entanglement entropy is relatively straightforward because it only involves tracing out one subsystem before subsequently diagonalizing the resultant density matrix to calculate its entropy.
As a result, bipartite systems, have long been the workhorses in quantum information processing~\cite{Masanes} and fundamental tests of quantum theory~\cite{Werner,Girolami}.
The natural extension of bipartite systems are multipartite systems, in which multiple subsystems must be traced out before diagonalizing the resultant density matrix~\cite{Coecke}, and which are central to many interesting tasks in quantum information~\cite{Toth,Ren}.

\begin{figure}[t]
\centerline{\includegraphics[width=\columnwidth]{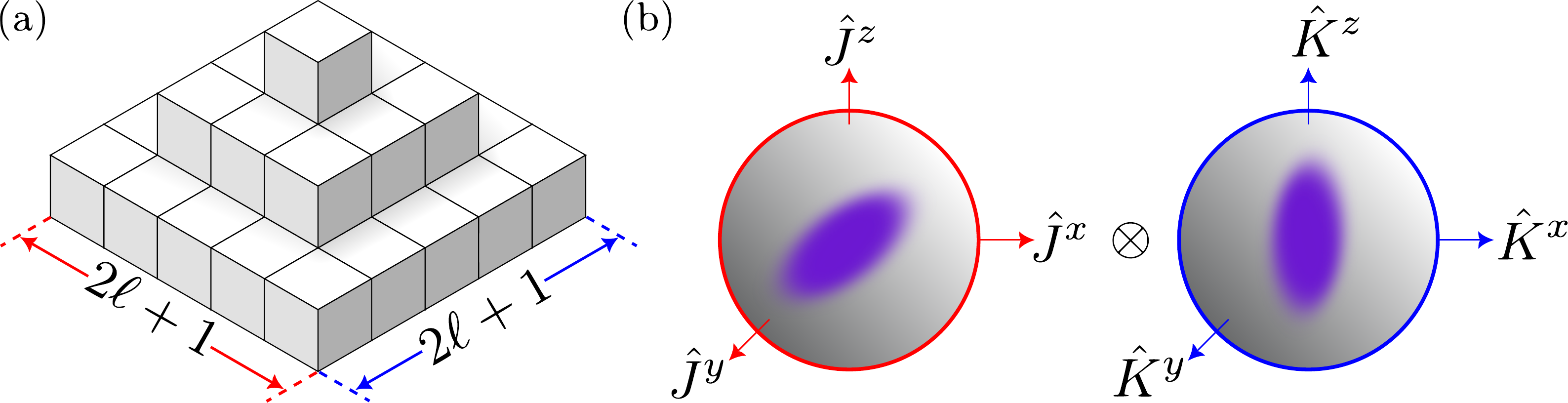}}
    \caption{
    (a) A cartoon of the pyramid structure of the $\SU{4}$ state space. 
    Each layer of the pyramid is a square of size $(2 \ell + 1)$-by-$(2 \ell + 1)$ for the $(2 \ell + 1)$-many levels of each degree of freedom.
    (b) The two Bloch spheres with dipole length $\ell$ for the $\SU{2}\otimes\SU{2}$ subgroup of $\SU{4}$. 
    Layer $\ell$ of the state space in (a) corresponds to many copies of this subgroup.
    }
    \label{fig:cartoon}
\end{figure}

Recently, there has been growing interest in physical processes that generate not only multipartite \textit{inter}particle entanglement, but also \textit{intra}particle entanglement amongst multiple degrees of freedom of a single particle or, more generally, entanglement between different degrees of freedom of different particles.
Borrowing terminology from quantum field theory, we will refer to entanglement between different degrees of freedom of many-body states as \emph{algebraic} entanglement~\cite{Balachandran,ReyesLega,Pontello} between degrees of freedom to distinguish it from multipartite entanglement between atoms regardless of the degrees of freedom.
A special case of this kind of entanglement is hybrid entanglement between a discrete degree of freedom and a continuous degree of freedom~\cite{Li2,VanLoock,Takeda,Andersen,Ulanov,Guccione,Shukla,Xu2}, but we can also use algebraic entanglement to encapsulate entanglement between two discrete (or two continuous) degrees of freedom in collective many-body states. 
These systems typically involve the internal energy levels of an atom as well as the external spatial degrees of freedom ~\cite{Wilson,Christopher_Wilsons_PaperAnd_Definitely_Not_Jarrods_Or_Johns,Gingrich,Jeske,Manzoni,Stav,Kale,Greve,Finger,Chelpanova,Barkhausen,Reilly} where the entanglement can be relevant for processes such as multi-parameter sensing~\cite{Christopher_Wilsons_PaperAnd_Definitely_Not_Jarrods_Or_Johns}, thermalization~\cite{kaufman2016}, and entropy removal via laser cooling~\cite{Metcalf,Corder,Bartolotta,Reilly4}.
Therefore, one would like to calculate the algebraic entanglement entropy between two or more degrees of freedom across the whole of the particle ensemble.
However, the process of unraveling these degrees of freedom by tracing one out is generally computationally expensive. 
For example, naively using the single-particle basis to trace out one degree of freedom scales exponentially with the number of particles which makes it intractable other than for a few particles.

In this paper, we describe how to utilize symmetries of a system and representation theory to calculate information theoretic quantities efficiently in polynomial complexity. 
We present a deep connection between the algebraic entanglement entropy amongst collections of particles with multiple degrees of freedom and the structure of Lie groups which describe these degrees of freedom.
We focus on the simplest case of an ensemble whose constituent particles have two degrees of freedom, such as spin and momentum, and both degrees of freedom only have two states.
In this case, we make use of the subgroup structure of $\SU{4}$ to trace out one of the degrees of freedom by finding an efficient direct mapping into a representative class of reduced density matrices which preserves the polynomial-scaling of the bosonic subspace of $\SU{4}$.
After tracing out one degree of freedom, we are left with a direct sum over density matrices describing the other degree of freedom using the irreducible representations (irreps) of $\SU{2}$.
As a consequence, the reduced density matrix in one of the subsystems takes on a block diagonal form which is again polynomial-scaled~\cite{Xu}.
We also provide a natural depiction of the state space into a pyramid structure, where each layer of the pyramid provides a representation of $\SU{2}\otimes\SU{2}$, i.e., two ``spin-ladders'' of equal lengths, as depicted in the cartoon of~\cref{fig:cartoon}.

Systems with this structure naturally appear in experiments where atoms are collectively coupled through a resonator, such as an optical cavity~\cite{PineiroOrioli,Wilson,Sundar,Silva,Reilly}.
The use of the $\SU{4}$ group structure allows one to reduce the scaling of computations from exponential $4^N$, to polynomial $\mathcal{O}(N^{3})$, where $N$ is the number of constituent particles.
This allows theoretical studies to perform exact simulations for $N \gg 1$ while keeping track of all interparticle and intraparticle correlations.
Moreover, the pyramid structure presented here and in Ref.~\cite{Xu} greatly reduces the task of both taking the trace and diagonalizing the density matrix in order to calculate quantities such as the quantum Fisher information~\cite{Liu3} and entanglement entropy~\cite{Horodecki}.
This is because these procedures can be performed in a block-by-block manner, the largest of which only scales as $\mathcal{O}(N^2)$, which is a massive reduction of numerical complexity compared to the single-particle basis. 

While we focus on the case $\SU{2} \otimes \SU{2} < \SU{4}$ here, where $<$ denotes a subgroup, our methodology can be extended to systems where each degree of freedom of the constituent particles possess a Lie group structure which can be decomposed into its irreps. 
Counterintuitively, we show that the entanglement entropy $S_E$ in these systems grows linearly with the number of constituent particles, $S_E = \mathcal{O}(N)$, rather than with the logarithm of the dimension of the polynomial-scaled Hilbert space, $S_E = \mathcal{O}(\ln(N))$, which is only possible due to the multiplicity (number of copies) of the irreps. 
This is a signature of complex particle-particle correlated physics that is a result of collective interactions in non-bosonic representations of the different degrees of freedom.
All of this leads to interesting, highly entangled physics which is straightforward to simulate in polynomial complexity with respect to $N$.

The paper is structured as follows.
We begin in Sec.~\ref{sec:IllustrativeExample} with an illustrative example of the $\SU{4}$ systems we consider in order to highlight the computational complexity of calculating algebraic entanglement entropy in the single-particle basis. 
We then turn to the symmetry properties and representation theory that allows us to reduce the computational complexity to polynomial-scaling in Sec.~\ref{sec:BosonicSU4}.
In Sec.~\ref{sec:entropy}, we present the full algorithm to calculate algebraic entanglement entropy by mapping the bosonic subspace of $\SU{4}$ to the irreps of $\SU{2}$.
In Sec.~\ref{sec:examples}, we provide a number of physical examples of both closed and open quantum systems where we apply our algebraic entanglement entropy algorithm.
We finish with some concluding remarks in Sec.~\ref{sec:Conclusion}.

\section{Illustrative Example} \label{sec:IllustrativeExample}
Before diving into the mathematical structure that we exploit to perform polynomial-scaled calculations of algebraic entanglement entropy, we first consider a simple physical example that can be solved analytically.
In this example, we can calculate the entanglement entropy by straightforward physical arguments.
However, we highlight that with more complicated $\SU{4}$ dynamics, direct methods are extremely computationally demanding without exploiting permutation symmetry and representation theory.

Consider a pulse from counterpropagating lasers applied to a cloud of $N$ atoms, driving a transition between two internal levels of the atoms, labeled $\ket{0}$ and $\ket{1}$ for ground and excited state, respectively. 
We assume that the atoms only have momentum along the lasers' axes, which we label the $z$-direction. 
By conservation of momentum, the atoms will experience a corresponding momentum impulse of $\pm \hbar k$ when absorbing or emitting photons from the lasers, where $k$ is the wavenumber of the photons.
During the time $t$ the pulse is applied for, the Hamiltonian of the system is given by
\begin{equation}
    \hat{H}_{\mathrm{CPL}} = \sum_{j=1}^N \frac{\hat{p}_j^2}{2m} + \frac{\hbar \Delta}{2} \hat{\sigma}^z_j + \hbar \Omega \cos( k \hat{z}_j ) \hat{\sigma}^x_j,
\end{equation}
where $\hat{z}_j$ is the position operator of atom $j$, $\hat{p}_j$ is the momentum operator, and $\hat{\sigma}_\mu^{(j)}$ is a Pauli operator with $\mu \in \{ x,y,z \}$.
Here, we have assumed that the lasers have the same Rabi frequency $\Omega$ and a fixed detuning $\Delta$ from the atomic transition frequency. 
When the energy of the photons is insufficient to drive the atoms to higher momentum states, as is the case in Refs.~\cite{Wilson,Christopher_Wilsons_PaperAnd_Definitely_Not_Jarrods_Or_Johns,Reilly}, one is able to truncate the atomic motion to just two momentum states, $\up_j \equiv \ket{+\hbar k / 2}_j$ and $\down_j \equiv \ket{- \hbar k / 2}_j$, assuming the atoms began with its population completely in these states.
In principle, there could be a momentum offset such as $\ket{p_0 \pm \hbar k/2}$ or the two momentum states could be separated by a multiple of $\hbar k$ and driven by multi-photon processes~\cite{Reilly3}, but we will only consider $\up_j$ and $\down_j$ here.
We can then trace out the remainder of the momentum basis such that the Hamiltonian becomes
\begin{equation} \label{H_E}
    \hat{H}_{\mathrm{IE}} = \sum_{j = 1}^N \frac{\hbar \Delta}{2} \hat{\sigma}^z_j + \frac{\hbar \Omega}{2} \hat{\sigma}^x_j \otimes \hat{s}^x_j,
\end{equation}
where we have dropped a constant kinetic energy shift of $N \hbar^2 k^2 / (8 m)$.

We can now study how the algebraic entanglement between the internal states and momentum states grows throughout the pulse duration. 
To do so, we assume the atoms are initially in the ground state with a momentum of $- \hbar k/2$: $\ket{\Psi (0)} = \bigotimes_{j = 1}^N (\ket{0}_j \otimes \down_j)$.
Since Eq.~\eqref{H_E} has no particle-particle interactions, we know the final state is separable atom-by-atom, i.e., $\ket{\Psi (t)} = \bigotimes_{j = 1}^N \ket{\psi (t)}_j$.
This observation will allow us to calculate the algebraic entanglement entropy simply. 
The state of atom $j$ may be solved for~\cite{Steck}
\begin{equation}
    \begin{aligned} \label{eq:analyticSol}
\ket{\psi(t)}_j = & e^{- \frac{i \Delta t}{2}} \left[ \cos(\frac{\tilde{\Omega} t}{2}) + \frac{i \Delta}{\tilde{\Omega}} \sin(\frac{\tilde{\Omega} t}{2}) \right] \ket{0}_j \otimes \down_j \\
&- i e^{- \frac{i \Delta t}{2}} \frac{\Omega}{\tilde{\Omega}} \sin(\frac{\tilde{\Omega} t}{2}) \ket{1}_j \otimes \up_j,
    \end{aligned}
\end{equation}
where $\tilde{\Omega} = \sqrt{\Omega^2 + \Delta^2}$ is the generalized Rabi frequency.
In the case that the drive is on resonance with the transition frequency, $\Delta = 0$, then at time $t_\mathrm{max} = \pi / (2 \Omega)$ we find the state  
\begin{equation}\label{eq:Psimax}
    \ket{\Psi_{\mathrm{max}}} = \bigotimes_{j = 1}^N \frac{1}{\sqrt{2}} (\ket{0}_j \otimes \up_j - i \ket{1}_j \otimes \down_j),
\end{equation}
which, notably, has maximum algebraic entanglement between atom $j$'s internal and momentum states.
We can see this by first tracing out either the internal states or the momentum states.
For notational clarity, we identify the two degrees of freedom by the labels $J$ and $K$ for the internal and external degrees of freedom, respectively.
The reduced density matrices are then $\hat{\rho}_{J}(t) = \Tr_{K}[\hat{\rho}(t)]$ and $\hat{\rho}_{K}(t) = \Tr_{J}[\hat{\rho}(t)]$ for the density matrix $\hat{\rho}(t) = \op{\Psi(t)}{\Psi(t)}$, where $\Tr_i$ denotes the partial trace over subsystem $i$. 
We can then calculate the entanglement entropy by finding the von Neumann entropy of the reduced density matrices,
\begin{equation} \label{eq:S_E}
    S_E [\hat{\rho}_{J} (t)] = - \Tr[ \hat{\rho}_{J}(t) \ln \hat{\rho}_{J}(t)] = S_E [\hat{\rho}_{K} (t)].
\end{equation}
The second equality holds for pure states by the monogamy of entanglement~\cite{Horodecki}.
It is straightforward to show that the state given by Eq.~\eqref{eq:Psimax} will have entanglement entropy scaling as 
\begin{equation} \label{eq:S_E_max}
    S_E [\hat{\rho}_{J,\mathrm{max}}] = N \ln 2,
\end{equation}
after noting that the reduced density matrix becomes the normalized $2^N$-dimensional identity operators after taking the partial trace,
\begin{equation}
    \hat{\rho}_{J,\mathrm{max}} = \Tr_K( \op{\psi_{\mathrm{max}}}{\psi_{\mathrm{max}}} ) = \frac{1}{2^N} \hat{\mathbb{I}}_{2^N}.
\end{equation}
Notably, this yields an algebraic entanglement entropy that scales linearly with $N$, rather than the natural log of $N$.
This linear scaling of the entanglement entropy is typically indicative of a Hilbert space which grows exponentially.

Note that the state Eq.~\eqref{eq:Psimax} has maximal algebraic entanglement while having no particle-particle entanglement. 
This means that the state would be extremely useful for quantum information science such as quantum teleportation between the two degrees of freedom, but would provide no quantum advantage beyond the standard quantum limit in quantum metrology.
In this case, the algebraic entanglement entropy is thus purely intraparticle entanglement between the degrees of freedom of each atom.
However, in more general systems, the calculation in Eq.~\eqref{eq:S_E} cannot differentiate this intraparticle entanglement from entanglement between, e.g., the $J$ degree of freedom of atom $i$ and the $K$ degree of freedom of atom $j$ ($i \neq j$). 
Moreover, the state with maximal hyperentanglement~\cite{Li3,Zeng}, $\ket{\Psi_{\mathrm{HE,max}}} = \ket{\Phi}_J \otimes \ket{\Xi}_K$ where $\ket{\Phi}_J$ and $\ket{\Xi}_K$ are symmetric $N$-particle Greenberger–Horne–Zeilinger (GHZ) states on the respective degree of freedom, has no algebraic entanglement between $J$ and $K$ under Eq.~\eqref{eq:S_E}. 
This is why we distinguish the algebraic entanglement calculated in Eq.~\eqref{eq:S_E} from intraparticle entanglement, multipartite entanglement between atoms, and hyperentanglement, while hybrid entanglement is a special case of algebraic entanglement with $J$ being discrete and $K$ being continuous or vice versa. 
Here, ``algebraic'' refers to the algebras associated with the symmetries of the respective degrees of freedom.
In the case above, we are calculating the entanglement between the $\mathfrak{su} (2)$ Lie algebra associated with $J$ and the $\mathfrak{su} (2)$ Lie algebra associated with $K$, as discussed further in the next section. 
We summarize the different classes of entanglement discussed above in Appendix~\ref{sec:EntClasses} (see Fig.~\ref{fig:VennDiag}).

In systems where the Hamiltonian (or jump operators) contain interparticle interactions, the exponential scaling of the Hilbert space would typically make the above analytical calculation of the algebraic entanglement entropy infeasible.
This is the case when considering the systems presented in, e.g.,  Refs.~\cite{Wilson,Reilly} in which the dynamics are solved using a second quantization formalism which abstracts away the single-particle dynamics.
On first pass, it appears that this makes calculating the algebraic entanglement entropy both analytically and numerically impossible for $N \gg 1$ because of the exponential-scaling of the single-particle basis.
Decomposing the state into single-particle dynamics is not possible, and therefore tracing out one of these degrees of freedom appears to require knowledge about individual particle-particle correlations, while diagonalizing the density matrix necessitates addressing the whole Hilbert space.
As a result, general calculations of $S_E$ are far more difficult.

In the following section, we will outline the state space needed to compute general dynamics of this form in order to show that the space of bosonic wavefunctions naturally admits an algebraic decomposition following from the structure of $\SU{2} \otimes \SU{2} < \SU{4}$.
This, we will show, allows the algebraic entanglement entropy to be calculated efficiently using representation theory in all cases that the constituent particles possess permutational symmetry.

\section{Bosonic Subspace of \texorpdfstring{$\SU{4}$}{SU(4)}} \label{sec:BosonicSU4}
We now turn to the general group structure that we focus on throughout this paper.
As in the previous example, we consider permutationally symmetric particles with two intraparticle degrees of freedom.
We also assume the constituent degrees of freedom of each particle are themselves two-level systems, such that they admit a $\SU{2} \otimes \SU{2} < \SU{4}$ structure.
This could therefore relate to atoms with a two-level electronic manifold and two momentum states~\cite{Wilson,Christopher_Wilsons_PaperAnd_Definitely_Not_Jarrods_Or_Johns,Reilly} or two relevant hyperfine states~\cite{Sundar}, molecules with two electronic and two vibrational states, two distinguishable spinor Bose-Einstein condensates in an optical lattice, or even an atom with two correlated spatial degrees of freedom.
We now give a brief overview of algebraic operators that generate the system's dynamics in these systems, describe how they connect to the Lie group $\SU{4}$ for permutation symmetric systems, and then introduce a state space that allows for the polynomial-scaled entanglement entropy algorithm presented in Sec.~\ref{sec:entropy}.

\subsection{\texorpdfstring{$\SU{4}$}{SU(4)} Operators}
The operator space we consider is given by 
\begin{equation} \label{su4Generators}
    \hat{G}_{\mu \nu} \equiv \sum_{j=1}^N \hat{\sigma}_\mu^{(j)} \otimes \hat{s}_\nu^{(j)},
\end{equation}
for $\mu, \nu \in \{ I, x, y, z, (+), (-) \}$ and Pauli matrices $\hat{\sigma}_\mu^{(j)}$ and $\hat{s}_\nu^{(j)}$ acting on the first or second degree of freedom of particle $j$. 
The operators are shown explicitly in Appendix~\ref{sec:singleParticleOps}. 
Additionally, $\hat{\sigma}^{(j)}_I = \hat{s}^{(j)}_I = \hat{\mathbb{I}}_2$ is the two-by-two identity operator on each space so that $\hat{G}_{II}$ is the identity operator on the whole system's vector space.
Considering just the Hermitian operators $\mu, \nu \in \{ I, x, y, z \}$, one can build a $15$-dimensional operator basis for the $\su{4}$ Lie algebra~\cite{Christopher_Wilsons_PaperAnd_Definitely_Not_Jarrods_Or_Johns} which, under exponentiation, generates $\SU{4}$.
These are simply linear combinations of the traditional Gell-Mann matrices~\cite{Georgi}. 
The two distinguishable degrees of freedom we care about will again be labeled $J$ and $K$, and correspond to the $\su{2}$ sub-algebras given by the operators $\hat{J}_\mu \equiv \hat{G}_{\mu I}/2$ and $\hat{K}_\nu \equiv \hat{G}_{I \nu}/2$, which are sums over all particles of the Pauli matrices $\mu, \nu \in \{ x, y, z \}$ on only one degree of freedom.
For the raising/lowering operators $\mu, \nu \in \{ (+), (-) \}$, we have the collective ladder operators $\hat{J}_{\pm} \equiv \hat{G}_{(\pm) I}$ and $\hat{K}_{\pm} \equiv \hat{G}_{I (\pm)}$.
Meanwhile, operators that are not in the respective $\mathfrak{su} (2)$ sub-algebras, $\hat{G}_{(\mu \neq I)(\nu \neq I)}$, represent $\SU{4}$ operators that generate algebraic entanglement between the $J$ and $K$ degrees of freedom (sub-algebras). 

\subsection{State Space}
The key to efficiently calculating the algebraic entanglement entropy in general is the symmetry underlying the state space.
In particular, we assume that the atoms evolve under a master equation of Lindblad form in which the Hamiltonian and jump operators are comprised entirely of $\SU{4}$ operators Eq.~\eqref{su4Generators} (see, e.g., Sec.~\ref{sec:examples}), which means that the Liouvillian superoperator of the dynamics commutes with the three Casimir operators of $\SU{4}$~\cite{Georgi}. 
This ensures that the particles remain in a permutationally symmetric subspace under the system's dynamics (assuming they begin in this irrep)~\cite{Bastin}, i.e. the state remains symmetric under the exchange of particle indices, whereupon calculating the reduced density matrices is greatly simplified.
This vector space is a basis of the bosonic irrep of the group $\SU{4}$ and, as we will see, contains many copies of the $\SU{2}$ subgroups for $J$ and $K$.
The total single-particle state space, labeled $\mh_\mathrm{tot}$, scales exponentially as $\dim(\mh_{\mathrm{tot}}) = 4^N$.
The use of permutation symmetry, however, reduces the size of the state space from exponential size to polynomial, $\dim(\mh_\mathrm{sym}) = (N + 1)(N + 2)(N + 3)/6 \sim N^3$~\cite{Xu}. 
Here, $\mh_\mathrm{sym}$ is the space of states of the form 
\begin{equation} \label{eq:SchwingerBoson}
    \ket{\alpha,\beta,\gamma,\delta} = \mathcal{S}\left(\ket{1,\upp}^{\otimes \alpha} \ket{1,\ddown}^{\otimes \beta} \ket{0,\upp}^{\otimes \gamma} \ket{0,\ddown}^{\otimes \delta}\right),
\end{equation}
where $\mathcal{S}$ is the symmetrizer and normalization dictates that $\alpha+\beta+\delta+\gamma = N$.
The method of Schwinger bosons~\cite{Mathur} is often used to describe dynamics on this space, whereupon each state is described as a mode with particles created and annihilated therein.
This allows one to solve the dynamics in a general second quantization picture, but yields no insight into the algebraic entanglement entropy between $J$ and $K$.
An equivalent and far more convenient basis can be found through the following observation.

Consider the case of two particles, $N = 2$. 
The state 
\begin{equation} \label{eq:SingletState}
    \frac{1}{2} \left( \ket{0}_1 \ket{1}_2 - \ket{1}_1 \ket{0}_2 \right) \otimes \left( \up_1 \down_2 - \down_1 \up_2 \right)
\end{equation}
is permutationally symmetric because permuting the particles permutes the label in both the spin and momentum basis.
In other words, the total state remains symmetric if both degrees of freedom are in an anti-symmetric singlet pair.
On the other hand, if one degree of freedom is in one of the symmetric triplet states,
\begin{equation} \label{eq:TripletStates}
    \begin{aligned}
& \ket{0}_1 \ket{0}_2, \quad \ket{1}_1 \ket{1}_2, \quad \frac{1}{\sqrt{2}} \left( \ket{0}_1 \ket{1}_2 + \ket{1}_1 \ket{0}_2 \right), \\
& \down_1 \down_2, \quad \up_1 \up_2, \quad \frac{1}{\sqrt{2}} \left( \down_1 \up_2 + \up_1 \down_2 \right),
    \end{aligned}
\end{equation}
then the other degree of freedom must also be in one of the triplet states for the total state to be in the symmetric $\SU{4}$ subspace of the two particles. 
Using the Young tableaux technique to find the irreps of $\mathrm{SU} (n)$~\cite{Georgi,Pfeifer,Eichmann}, this argument can be extended to the full ensemble of particles so that both degrees of freedom must have the same number of singlet pairs, i.e., be in the same irrep of the $\SU{2}$ subgroups, for the total state to be in the symmetric $\SU{4}$ subspace~\cite{Xu}.
In simpler terms, the anti-symmetry of particle exchange in each degree of freedom has to be exactly the same in order to result in a net symmetry under particle exchange.

The states in Eqs.~\eqref{eq:SingletState} and~\eqref{eq:TripletStates} could equivalently be labeled by the eigenvalues of the collective $z$-operators and Casimir operators of the $\mathrm{SU} (2)$ subgroups,
\begin{equation}
    \begin{aligned}
\hat{J}_z &= \frac{\hat{G}_{z I}}{2},  & \hat{J}^2 = \sum_{\mu = x,y,z} \left( \hat{J}_{\mu} \right)^2 = \sum_{\mu = x,y,z} \left( \frac{\hat{G}_{\mu I}}{2} \right)^2, \\
\hat{K}_z &= \frac{\hat{G}_{I z}}{2}, & \hat{K}^2 = \sum_{\nu = x,y,z} \left( \hat{K}_{\nu} \right)^2 = \sum_{\nu = x,y,z} \left( \frac{\hat{G}_{I \nu}}{2} \right)^2.
    \end{aligned}
\end{equation}
The Dicke states $\ket{j,m_j}$ and $\ket{k,m_k}$ are then eigenvectors of these operators with eigenvalues given by $\hat{J}_z \ket{j,m_j} = m_j \ket{j,m_j}$, $\hat{J}^2 \ket{j,m_j}=j(j+1) \ket{j,m_j}$, $\hat{K}_z \ket{k,m_k} = m_k \ket{k,m_k}$, and $\hat{K}^2 \ket{k,m_k}=k(k+1) \ket{k,m_k}$.
The two-particle triplet states are then 
\begin{equation}
    \begin{aligned}
& \ket{j = 1, m_j = 1}, \quad \ket{j = 1, m_j = 0}, \quad \ket{j = 1, m_j = -1}, \\
& \ket{k = 1, m_k = 1}, \quad \ket{k = 1, m_k = 0}, \quad \ket{k = 1, m_k = -1},
    \end{aligned}
\end{equation}
while the singlet states are
\begin{equation}
    \ket{j = 0, m_j = 0}, \quad \ket{k = 0, m_k = 0}.
\end{equation}
This leads to the important observation that, so long as both degrees of freedom have the same Casimir eigenvalue $j = k$, the overall wavefunction will remain permutationally symmetric.
Extending this to the total ensemble of particles~\footnote{In terms of Young tableaux, the irreps of both $\mathrm{SU} (2)$ subgroups must have $(N/2 - J)$ many boxes in the second row indicating singlet pairs~\cite{Georgi,Pfeifer,Eichmann}.}, we can thus label an orthonormal basis of the total bosonic representation of $\SU{4}$ as $\ket{\ell, m_j, m_k}$, where
\begin{equation} \label{eq:state}
    \hat{J}^2 \ket{\ell,m_j,m_k} = \hat{K}^2 \ket{\ell,m_j,m_k} = \ell(\ell+1) \ket{\ell,m_j,m_k}.
\end{equation}
This new basis is equivalent to the $\ket{\alpha,\beta,\gamma,\delta}$ basis, where we technically only need to specify three quantum numbers $\alpha,\beta,\gamma$ to fully specify a state because $\delta = N - \alpha - \beta - \gamma$.
The two bases uniquely map into each other through $m_j = ( \alpha + \beta - \gamma - \delta )/2$, and $m_k = ( \alpha - \beta + \gamma - \delta )/2 $, while $\hat{J}^2$ and $\hat{K}^2$ are not diagonal in the Schwinger boson basis and so $\ell$ is found through explicit evaluation.
This is related to an important point: in general, $\ket{\ell,m_j,m_k} \neq \ket{j = \ell,m_j} \otimes \ket{k=\ell,m_k}$.
We know this is not the case because the state given in Eq.~\eqref{eq:Psimax} may be fully described by the permutationally symmetric basis states, but it is clearly not of this form.
Lastly, we can do an accounting of all the basis states.
The quantum number $\ell$ can range from $\ell=N/2$ down to $\ell=1/2$ or $0$ for odd or even $N$, respectively, in steps of one: $\ell \in \{ N / 2, N / 2 - 1, \ldots, 0 \; \mathrm{or} \; 1 / 2 \}$.
Each $\ell$ has $2\ell+1$ many values of $m_j$ and $m_k$, with $m_j, m_k \in \{ +\ell, +\ell-1,...,-\ell\}$.
This sum then gives the correct total number of states,
\begin{equation}
    \sum_{\ell}^{N/2} (2 \ell+1)^2 = (N + 1)(N + 2)(N + 3)/6.
\end{equation}

\subsection{\texorpdfstring{Irreducible Representations of $\SU{2}$}{SU(2)} Subgroups} \label{sec:irreps}
As stated above, the basis states $\ket{\ell,m_j,m_k}$ are not trivially separable into wavefunctions of the $J$ and $K$ degrees of freedom.
This is because each value $\ell$ corresponds to many different configurations of the atoms.
Consider, for example, the states $\ket{\ell=\frac{3}{2},m_j=\frac{1}{2},m_k=\frac{1}{2}}$ and $\ket{\ell=\frac{1}{2},m_j=\frac{1}{2},m_k=\frac{1}{2}}$ for $N = 3$.
For $\ell = 3 / 2$, we have
\begin{equation}
    \begin{aligned}
\ket{\frac{3}{2},\frac{1}{2},\frac{1}{2}} = & \frac{1}{3} \left( \ket{110}_{\mathrm{s}} + \ket{101}_{\mathrm{s}} + \ket{011}_{\mathrm{s}} \right) \\
& \otimes \left( \ket{\upp \upp \ddown}_{\mathrm{s}} + \ket{\upp \ddown \upp}_{\mathrm{s}} + \ket{\ddown \upp \upp}_{\mathrm{s}} \right),
    \end{aligned}
\end{equation}
where the states on the right-hand-side of the equation are in the single-particle basis with the notation $\ket{i j k}_{\mathrm{s}} = \ket{i}_1 \ket{j}_2 \ket{k}_3$. 
The $\ell = 1/2$ state is more complicated,
\begin{equation} \label{eq:exState}
    \begin{aligned}
\ket{\frac{1}{2},\frac{1}{2},\frac{1}{2}} &= \frac{1}{3\sqrt{2}} \left( \ket{110}_{\mathrm{s}} + \mathrm{e}^{i 2\pi/3} \ket{101}_{\mathrm{s}} + \mathrm{e}^{i 4 \pi/3} \ket{011}_{\mathrm{s}} \right) \\
& \otimes \left( \ket{\upp \upp \ddown}_{\mathrm{s}} + \mathrm{e}^{i 2\pi/3} \ket{\upp \ddown \upp}_{\mathrm{s}} + \mathrm{e}^{i 4 \pi/3} \ket{\ddown \upp \upp}_{\mathrm{s}} \right) \\
& + \frac{1}{3\sqrt{2}} \left( -\ket{110}_{\mathrm{s}} + \mathrm{e}^{i \pi/3} \ket{101}_{\mathrm{s}} + \mathrm{e}^{i 5 \pi/3} \ket{011}_{\mathrm{s}} \right) \\
& \otimes \left( - \ket{\upp \upp \ddown}_{\mathrm{s}} + \mathrm{e}^{i \pi/3} \ket{\upp \ddown \upp}_{\mathrm{s}} + \mathrm{e}^{i 5 \pi/3} \ket{\ddown \upp \upp}_{\mathrm{s}} \right). 
    \end{aligned}
\end{equation}
We therefore see that the state $\ket{\ell=\frac{3}{2},m_j=\frac{1}{2},m_k=\frac{1}{2}}$ has no algebraic entanglement between the two degrees of freedom whereas $\ket{\ell=\frac{1}{2},m_j=\frac{1}{2},m_k=\frac{1}{2}}$ has an algebraic entanglement entropy of $1$ bit. 
We emphasize again that both states are still overall permutationally symmetric.
This pattern will hold as $N$ increases.
In general, the $N$-particle bosonic irrep of $\SU{4}$ has every $\SU{2} \otimes \SU{2}$ representation from $\ell=N/2$ down to $\ell=1/2$ or $0$ for $N$ being odd or even, respectively.
Upon tracing out one degree of freedom, each $\SU{2}$ irrep is identified by a specific value of $\ell$ and will have a multiplicity of $d^\ell_N$ given by~\cite{Mandel}
\begin{equation}
    d^\ell_N = \frac{N! (2 \ell +1)}{(\frac{N}{2} + \ell + 1)! \ (\frac{N}{2} - \ell)!},
\end{equation}
which represents the number of ``copies'' of the state in the single-particle basis that only differ by a permutation of atom indices (i.e., they are states with different standard Young tableaux but the same semistandard Young tableau~\footnote{For a given partition (Young diagram), standard Young tableaux label basis states of the irreps of the permutation group $\mathrm{S}_N$, while semistandard Young tableaux (a.k.a. standard Weyl tableaux) encode weight vectors (collective states) of the corresponding irrep of $\mathrm{SU} (n)$. The connection between the two is provided by the Schur-Weyl duality. See Ref.~\cite{Bastin} for more.}), and thus evolve identically under a permutationally symmetric Hamiltonian. 
Note that $d^{N/2}_N = 1$, as expected for the bosonic subspace.

We can now use the $d^\ell_N$-fold multiplicity of each $\ell$-irrep of $\SU{2}$ to establish the following result.
If we trace out one of the degrees of freedom, say $K$, each basis state $\ket{\ell,m_j,m_k}$ is mapped to a density matrix with support over each of the $d^\ell_N$-many orthogonal copies of the $\SU{2}$ irrep.
Stated in an equation, this means that
\begin{equation}
    \begin{aligned} \label{eq:JBasisTrace}
\hat{\rho}_{\ell,m_j,m_j'} &= \Tr_K\left( \op{\ell,m_j,m_k}{\ell,m_j',m_k} \right) \\
&= \frac{1}{d^\ell_N} \sum_{i=1}^{d^\ell_N} \op{\ell,m_j,i}{\ell,m_j',i},
    \end{aligned}
\end{equation}
where $i$ is an extra quantum number (from outside the group) needed to distinguish each copy~\cite{Shankar}.
Notably, there are no coherences between different irreps of $\mathrm{SU} (2)$ in the reduced density matrix, and no coherences between different copies. 
For example, tracing out the $K$ degree of freedom of~\cref{eq:exState} would lead to $i=1$ being the outer product of the state in the first line and $i=2$ being the outer product of the state in the third line with no coherence between them.
Similarly, if we were to trace over the $J$ degree of freedom, we obtain
\begin{equation} 
    \begin{aligned} 
\hat{\rho}_{\ell,m_k,m_k'} &\equiv \Tr_J\left( \op{\ell,m_j,m_k}{\ell,m_j,m_k'} \right) \\
&= \frac{1}{d^\ell_N} \sum_{i=1}^{d^\ell_N} \op{\ell,m_k,i}{\ell,m_k',i}.
    \end{aligned}
\end{equation}
With this, to solve for the algebraic entanglement entropy in general, we just have to decompose it into the constituent density matrices on the $J$ and $K$ degrees of freedom.

\section{An Algorithm for Algebraic Entanglement Entropy} \label{sec:entropy}
With the structure of the state space in place and the multiplicity of the relevant subgroups worked out, we can now describe the structure of our algorithm for calculating algebraic entanglement entropy efficiently.
It is conceptually the same for pure versus mixed states, but we will describe them separately.
For pure states, the state vector can be described in $\mathcal{O}(N^3)$ coefficients, while mixed states require $\mathcal{O}(N^6)$ many coefficients to describe the density matrix.
Of course, the use of sparse matrices can reduce the numerical complexity drastically for most systems.

\begin{figure}[h]
\centerline{\includegraphics[width=\columnwidth]{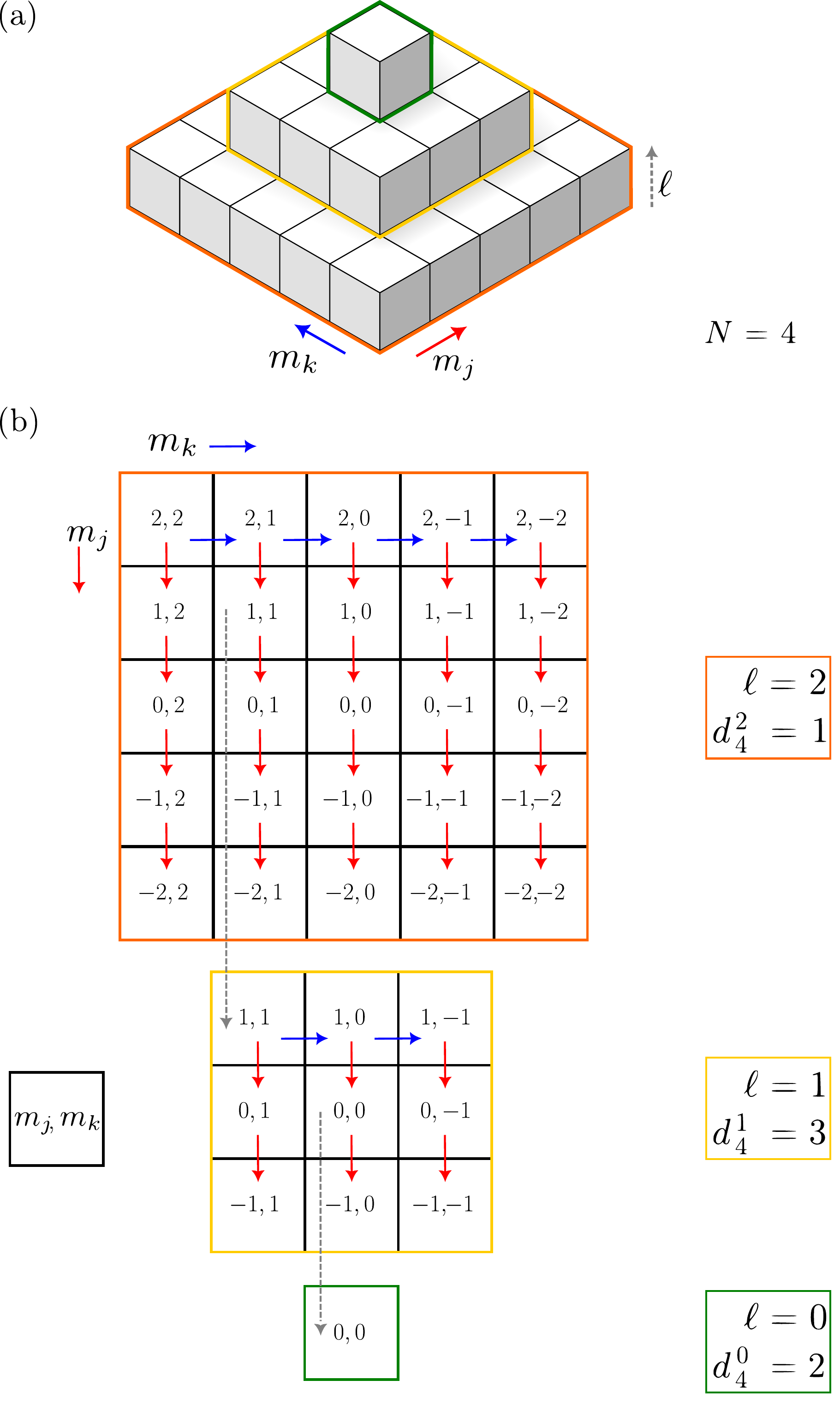}}
    \caption{
    (a) The state pyramid for $N=4$.
    (b) The layers of the state pyramid, where each $\ell$ corresponding to one of the irreducible representations of $\SU{2} \otimes \SU{2}$.
    Red arrows indicate an application of $\hat{J}_-$, blue arrows indicate an application of $\hat{K}_-$, and gray arrows indicate a Gram-Schmidt orthogonalization step. 
    The $\ell$ label and the multiplicity $d^{\ell}_N$ are given to the right of each layer.
    }
    \label{fig:Pyramid}
\end{figure}

It is helpful to imagine the states organized onto a three-dimensional grid, where each row is labeled by $m_j$, while each column is labeled by $m_k$, and where the third layer is labeled by $\ell$. 
This organizes the states into a pyramid structure shown in Fig.~\ref{fig:Pyramid}(a), where $\SU{2}\otimes\SU{2}$ dynamics don't mix the layers of the pyramid (don't change $\ell$), but $\SU{4}$ dynamics in general can couple states on different layers.
We show a picture of this for $N=4$ in~\cref{fig:Pyramid}(a).
The six generators of $\SU{2} \otimes \SU{2}$, given by
$\hat{G}_{\mu I}$ and $\hat{G}_{I \mu}$ for $\mu \in \{ x,y,z \}$,
don't move a state between layers of this pyramid. 
The remaining $\SU{4}$ operators, $\hat{G}_{\mu \nu}$ for $\mu, \nu \in \{ x,y,z \}$, can change $\ell$ and therefore generate algebraic entanglement.
This conclusion is equally clear from the definitions of~\cref{su4Generators}, and is important for the steps of the algorithm.
Moreover, our observation in the previous section about the multiplicity of the different $\ell$ layers leads to an important conclusion; basis states on layers $\ell \neq N / 2$ are automatically endowed (when $N > 2$) with algebraic entanglement between $J$ and $K$ through the use of representation theory.
More concretely, the multiplicity factor accounts for the compression of a linear superposition over copies of a given irrep in the basis of our simulation, e.g., Eq.~\eqref{eq:JBasisTrace}.
This fact is the sole reason we can calculate a linearly scaled entanglement entropy, $S_E \sim N$, using a polynomial-scaled Hilbert space.

\subsection{Pure State} \label{sec:pureEnt}
We begin with the case that the system undergoes unitary dynamics under a Hamiltonian composed solely of the $\SU{4}$ operators of Eq.~\eqref{su4Generators} such that it remains in a pure state $\ket{\psi}$. 
To calculate the algebraic entanglement entropy, one first has to find the state coefficients in the $\ket{\ell,m_j,m_k}$ basis to take advantage of the innate symmetry.
For each $\ket{\ell,m_j,m_k}$, we need to calculate the coefficients
\begin{equation} \label{eq:PureCoeff}
    p_{\ell,m_j,m_k} = \ip{\ell,m_j,m_k}{\psi}.
\end{equation}
Of course, one could simply simulate the state dynamics in this basis, but since the Schwinger boson basis Eq.~\eqref{eq:SchwingerBoson} is typically used, we will assume the state dynamics $\ket{\psi (t)}$ are simulated in the Schwinger boson basis and one wants to calculate the $p_{\ell,m_j,m_k}$ coefficients as the state evolves. 
To do so algorithmically, we start with the state coefficients
\begin{equation}
    \ket{\psi} = c_{\alpha,\beta,\gamma,\delta} \ket{\alpha,\beta,\gamma,\delta}.
\end{equation}
In this basis, we can first identify the state $\ket{\ell=N/2,m_j=N/2,m_k=N/2}=\ket{N,0,0,0}$. 
In principle, one could start in any preferred basis (e.g., Ref.~\cite{Reilly}) so long as one can identify the state $\ket{\ell=N/2,m_j=N/2,m_k=N/2}$.
We know that $N$ applications of $\hat{J}_-$ on this state will produce $\ket{\ell=N/2,m_j=-N/2,m_k=N/2}$, while $N$ applications of $\hat{K}_-$ will produce $\ket{\ell=N/2,m_j=N/2,m_k=-N/2}$.
As a result, we can iterate over the $\ell=N/2$ states using $\hat{J}_-$ and $\hat{K}_-$, whereupon
\begin{equation}
    p_{\frac{N}{2},m_j,m_k} = \left \langle \frac{N}{2},\frac{N}{2},\frac{N}{2} \left \lvert \frac{(\hat{J}_+)^{\frac{N}{2} - m_j} (\hat{K}_+)^{\frac{N}{2} - m_k} }{\mathcal{N}_{\frac{N}{2},m_j,m_k}} \right \rvert \psi \right \rangle,
\end{equation}
where $\mathcal{N}_{\ell,m_j,m_k}$ is a normalization factor after the raising operators are applied.

In order to generate the states of $\ell=\frac{N}{2} - 1$, we use the state $\ket{N/2,N/2-1,N/2-1}$ and Gram-Schmidt orthogonalization~\cite{GriffithsText,Bjorck} to find $\ket{N/2 - 1,N/2 - 1,N/2 - 1}$ such that
\begin{equation}
    \ip{\frac{N}{2},\frac{N}{2}-1,\frac{N}{2}-1}{\frac{N}{2} - 1,\frac{N}{2} - 1,\frac{N}{2} - 1} = 0.
\end{equation}
We can then generate the next ``layer'' of state coefficients using
\begin{equation}
    p_{\frac{N}{2}-1,m_j,m_k} = \left \langle \frac{N}{2} - 1,\ell,\ell \left \lvert \frac{(\hat{J}_+)^{\ell - m_j} (\hat{K}_+)^{\ell - m_k} }{\mathcal{N}_{\ell,m_j,m_k}} \right \rvert \psi \right \rangle.
\end{equation}
This process is repeated for each value of $\ell$, where the $n^\mathrm{th}$ layer is found via Gram-Schmidt orthogonalization using the $n-1$ states sharing the same $m_j$ and $m_k$, but differing in $\ell$.
This iterative process to find all the state coefficients $p_{\ell,m_j,m_k}$ is shown pictorially in~\cref{fig:Pyramid}(b), where the red arrows depict $\hat{J}_-$ operations, the blue arrows depict $\hat{K}_-$ operations, and the gray arrows depict the Gram-Schmidt orthogonalization steps.

It is now a straightforward process to trace out one of the degrees of freedom, say $K$. 
We use Eq.~\eqref{eq:JBasisTrace} to find
\begin{equation} \label{eq:PureRhoJ}
    \begin{aligned}
\hat{\rho}_J &= \Tr_K( \op{\psi} ) \\
&= \sum_{\ell}^{N/2} \left[ \sum_{m_j,m_j',m_k=-\ell}^\ell p_{\ell,m_j,m_k} p^*_{\ell,m_j',m_k}
\ \hat{\rho}_{\ell,m_j,m_j'} \right],
    \end{aligned}
\end{equation}
where the sum on $\ell$ starts at $0$ or $1/2$ for even or odd $N$, respectively.
We can use this form because $\hat{\rho}_J$ will never support coherences between layers of differing $\ell$.
A similar equation for $\hat{\rho}_K$ is found when tracing out the other degree of freedom. 
Now, we may simply diagonalize each layer one-by-one to find the eigenvalues of $\hat{\rho}_J$.
We do this by constructing the matrix
\begin{equation}
    \mathbf{M}^{(\ell)} = \sum_{m_j,m_j',m_k = -\ell}^\ell p_{\ell,m_j,m_k}p_{\ell,m_j',m_k} \ \mathbf{E}^{(\ell)}_{m_j + \ell, m_j' + \ell},
\end{equation}
where $\mathbf{E}^{(\ell)}_{\mu\nu}$ is a $(2 \ell + 1)\times(2 \ell+1)$ matrix with a $1$ in the $\mu^\mathrm{th}$ row and $\nu^\mathrm{th}$ column, and zeros everywhere else. 
Therefore, $\mathbf{M}^{(\ell)}$ is the matrix whose elements are populated by each $p_{\ell,m_j,m_k} p^*_{\ell,m_j',m_k}$ coefficient.
Diagonalizing this matrix yields the eigenvalues for the $\ell^\mathrm{th}$ component of $\hat{\rho}_J$.
Defining $\lambda^{(\ell)}_i$ to be the $i^\mathrm{th}$ eigenvalue of $\mathbf{M}^{(\ell)}$, we must remember that there are $d^{\ell}_N$-many copies of the eigenvalue $\lambda^{\ell}_i / d^{\ell}_N$ in the full spectrum of $\hat{\rho}_J$ due to the multiplicity of the given irrep.

Lastly, the algebraic entanglement entropy of $\hat{\rho}_J$ may be directly calculated by summing over the eigenvalues of each $\mathbf{M}^{(\ell)}$ and including the proper $d^\ell_N$-fold degeneracy of these eigenvalues,
\begin{equation}
    \begin{aligned} \label{eq:ent}
S_E [\hat{\rho}_J] &= \Tr_J [ \hat{\rho}_J \ln(\hat{\rho}_J) ] \\
&= - \sum_{\ell}^{N/2} \left[ \sum_{i = 0}^{2 \ell + 1} \lambda_i^{(\ell)} \ln\left( \frac{\lambda^{(\ell)}_i}{d^\ell_N} \right) \right], 
    \end{aligned}
\end{equation}
where we take $0 \times \ln( 0 ) = 0$ for any zero eigenvalues.
We note that each $\lambda_i^{(\ell)}/d^\ell_N$ appears $d^\ell_N$-many times, which is why the first occurrence of $\lambda_i^{(\ell)}$ in Eq.~\eqref{eq:ent} doesn't have an accompanying $d^\ell_N$.
We again emphasize the role of the multiplicity of the $\SU{2}$ irreps to accurately calculate the algebraic entanglement entropy.
For example, consider $\ket{\psi} = \ket{1/2,1/2,1/2}$ for $N = 3$ which we wrote in the single-particle basis in Eq.~\eqref{eq:exState}. 
Since the state is purely in a single basis state, we have a single nonzero eigenvalue $\lambda_1^{(1/2)} = 1$ which one would often think indicates zero entropy.
However, the multiplicity of $d_3^{1/2} = 2$ in Eq.~\eqref{eq:ent} corrects this and gives $S_E[\hat{\rho}_J] = \ln(2) = 1 \; \mathrm{bit}$, which agrees with what we found in Eq.~\eqref{eq:exState}.

The pure state algorithm presented above is summarized in pseudocode in Algorithm~\ref{alg:EE}.
There, we show algorithmically how to start with a state $\ket{\psi}$ in the Schwinger boson basis for a given $N$ and calculate the entanglement entropy for $\hat{\rho}_J$ and $\hat{\rho}_K$.

\begin{algorithm}[ht]
\label{alg:EE}
\SetAlgoLined
Begin with a state $\ket{\psi}$ in $\ket{\alpha, \beta, \gamma, \delta}$ basis\;
Initialize $\hat{J}_-$ and $\hat{K}_-$ in $\ket{\alpha, \beta, \gamma, \delta}$ basis\;
$\ell_{\mathrm{min}} \leftarrow \mathrm{mod}[N,2] / 2$\;
$S_J \leftarrow 0$\;
$S_K \leftarrow 0$\;
\For{$\ell = \frac{N}{2}:-1:\ell_{\mathrm{min}}$}{
     $\mathbf{p}_{\ell} \leftarrow \mathrm{zeros}(2 \ell + 1, 2 \ell + 1)$\;
     \eIf{$\ell = \frac{N}{2}$}{$\ket{s'} \leftarrow \ket{N,0,0,0}$\;}{Gram-Schmidt orthogonalization for $\ket{s'}$\;}
     \For{$j = 1:2 \ell + 1$}{
        $\ket{s} \leftarrow \ket{s'}$\;
        $\mathbf{p}_{\ell}(j,1) \leftarrow \ip{s}{\psi}$\;
        \For{$k = 2:2 \ell + 1$}{
            $\ket{s} \leftarrow \hat{K}_- \ket{s}$\;
            Normalize $\ket{s}$\;
            $\mathbf{p}_{\ell}(j,k) \leftarrow \ip{s}{\psi}$\;
        }
        $\ket{s'} \leftarrow \hat{J}_- \ket{s'}$\;
        Normalize $\ket{s'}$\;
     }
     $d_N^{\ell} \leftarrow N! (2 \ell + 1) / [(N / 2 + \ell + 1)! (N / 2 - \ell)!]$\;
     $\mathbf{M}^{(\ell)}_J(j,j') \leftarrow \sum_{k = 0}^{2 \ell + 1} \mathbf{p}_{\ell}(j,k)\mathbf{p}_{\ell}^*(j',k)$\;
     $\mathbf{M}^{(\ell)}_K(k,k') \leftarrow \sum_{j = 0}^{2 \ell + 1} \mathbf{p}_{\ell}(j,k)\mathbf{p}_{\ell}^*(j,k')$\;
     $\lambda_J \leftarrow \mathrm{eigenvalues}(\mathbf{M}^{(\ell)}_J)$\;
     $\lambda_K \leftarrow \mathrm{eigenvalues}(\mathbf{M}^{(\ell)}_K)$\;
     $S_J \leftarrow S_J - \sum_{j = 0}^{2 \ell + 1} \lambda_J(j) \ln(\lambda_J(j) / d^{\ell}_N)$\;
     $S_K \leftarrow S_K - \sum_{k = 0}^{2 \ell + 1} \lambda_K(k) \ln(\lambda_K(k) / d^{\ell}_N)$\;
}
\caption{The algorithm for finding the algebraic entanglement entropy of pure state $\ket{\psi}$.
The algorithm loops over each ``layer'' of states for one value of $\ell$ ranging from $N/2$ down to either $1/2$ or $0$.
The starting state for layer $\ell$ is found by Gram-Schmidt orthogonalization using all previous states that have $m_j = m_k = \ell$.
From this starting state, all other states are reached in the layer by repeated application of $\hat{J}_-$ and $\hat{K}_-$.
These states are used to calculate the state coefficients of $\ket{\psi}$, and subsequently construct the matrix $\mathbf{M}$.
Lastly, the eigenvalues of this matrix and the multiplicity $d^{\ell}_N$ are used to calculate the entanglement entropy.}
\end{algorithm}

\subsection{Mixed State} \label{sec:Mixedent}
We now allow the system to also undergo dissipative dynamics under a master equation of Lindblad form with jump operators comprised entirely of $\SU{4}$ operators, and so we must generalize the above procedure for mixed quantum states.
In general, this requires $\mathcal{O}[ \dim(\mathcal{H})^2 ] = \mathcal{O}(N^6)$ many coefficients to describe the density matrix $\hat{\rho}$.
However, the algorithm is effectively the same for calculating $\hat{\rho}_J$ and $\hat{\rho}_K$.
The only non-trivial change is that we will have coefficients that depend on two copies of $\ell, m_j,$ and $m_k$:
\begin{equation}
    P_{\ell,m_j,m_k}^{\ell',m_j',m_k'} = \bra{\ell,m_j,m_k} \hat{\rho} \ket{\ell', m_j', m_k'}.
\end{equation}
If $\hat{\rho}$ were a pure state, it is straightforward to see that $P_{\ell,m_j,m_k}^{\ell',m_j',m_k'} =  p_{\ell,m_j,m_k} p^*_{\ell',m_j',m_k'}$ from~\cref{eq:PureCoeff}.
This means that, for general mixed states, one must iterate over two sets of states, $\ket{\ell,m_j,m_k}$ and $\ket{\ell',m_j',m_k'}$, to calculate all values of $P_{\ell,m_j,m_k}^{\ell',m_j',m_k'}$. 

Once these coefficients are calculated, we can express any general density matrix, $\hat{\rho}$, using the pyramid structure:
\begin{equation}
\begin{aligned}
& \hat{\rho} = \sum_{\ell,\ell'}^{N/2} \sum_{\mathbf{m}} P_{\ell,m_j,m_k}^{\ell',m_j',m_k'} \op{\ell, m_j, m_k}{\ell', m_j', m_k'}.
\end{aligned}
\end{equation}
where $\mathbf{m} = m_j,m_k,m_j',m_k'$ for legibility, and the sums over $m_j$, $m_k$ run from $-\ell$ to $+\ell$, while the sums over $m_j'$, $m_k'$ run from $-\ell'$ to $+\ell'$.
The rest of the procedure is the same as for pure states because we know $\hat{\rho}_J$ and $\hat{\rho}_K$ will have no coherence between differing values of $\ell$ and $\ell'$.
Therefore, we have
\begin{equation}
    \begin{aligned}
\hat{\rho}_J &= \Tr_K( \hat{\rho} ) \\
&= \sum_{\ell}^{N/2} \left[ \sum_{m_j,m_j',m_k} P_{\ell,m_j,m_k}^{\ell,m_j',m_k}
\ \hat{\rho}_{\ell,m_j,m_j'} \right],
    \end{aligned}
\end{equation}
where $m_j,m_j',m_k$ are summed from $-\ell$ to $\ell$.
This agrees with the pure state case~\cref{eq:JBasisTrace}, and we can find a similar equation for $\hat{\rho}_K$.

Once again, we diagonalize each layer for distinct $\ell$ one-by-one to find the eigenvalues of $\hat{\rho}_J$.
We do this by constructing the matrix
\begin{equation}
    \mathbf{M}^{(\ell)} = \sum_{m_j,m_j',m_k = -\ell}^\ell P_{\ell,m_j,m_k}^{\ell,m_j',m_k} \ \mathbf{E}^{(\ell)}_{m_j + \ell, m_j' + \ell}.
\end{equation}
Diagonalizing this matrix yields the eigenvalues for the $\ell^\mathrm{th}$ component of $\hat{\rho}_J$.
The algebraic entanglement entropy of $\hat{\rho}_J$ may then be directly calculated in an equivalent manner to~\cref{eq:ent}.
For mixed states, however, we note that $S_E[\hat{\rho}_J] \neq S_E[\hat{\rho}_K]$ as the system may also be entangled to an external environment. 
The algorithm for mixed states presented above is summarized in pseudocode in Algorithm~\ref{alg:EEmixed}.

\begin{algorithm}[ht]
\label{alg:EEmixed}
\SetAlgoLined
Begin with a state $\hat{\rho}$ in $\ket{\alpha, \beta, \gamma, \delta}$ basis\;
Initialize $\hat{J}_-$ and $\hat{K}_-$ in $\ket{\alpha, \beta, \gamma, \delta}$ basis\;
$\ell_{\mathrm{min}} \leftarrow \mathrm{mod}[N,2] / 2$\;
$S_J \leftarrow 0$\;
\For{$\ell = \frac{N}{2}:-1:\ell_\mathrm{min}$}{
     $s_a \leftarrow$ Empty $(2 \ell + 1) \cross (2 \ell + 1)$ array\;
     \eIf{$\ell = \frac{N}{2}$}{$\ket{s'} \leftarrow \ket{N,0,0,0}$\;}{Gram-Schmidt orthogonalization for $\ket{s'}$\;}
     \For{$j = 1:2 \ell + 1$}{
        $\ket{s} \leftarrow \ket{s'}$\;
        $s_a(j,1) = \ket{s}$\;
        \For{$k = 2:2 \ell + 1$}{
            $\ket{s} \leftarrow \hat{K}_- \ket{s}$\;
            Normalize $\ket{s}$\;
            $s_a(j,k) = \ket{s}$\;
        }
        $\ket{s'} \leftarrow \hat{J}_- \ket{s'}$\;
        Normalize $\ket{s'}$\;
     }
     $P \leftarrow \mathrm{zeros}(2 \ell + 1, 2 \ell + 1, 2 \ell + 1)$\;
     \For{$j,j',k = 1:2 \ell + 1$}{
        $P(j,j',k) \leftarrow s_a(j,k)^{\dagger} \hat{\rho} s_a(j',k)$\;
     }
     $\mathbf{M}_J^{(\ell)} \leftarrow \mathrm{zeros}(2 \ell + 1, 2 \ell + 1)$\;
     \For{$j,j' = 1:2 \ell + 1$}{
        $\mathbf{M}_J^{(\ell)}(j,j') \leftarrow \sum_{k = 0}^{2 \ell + 1} P(j,j',k) $\;
     }
     $\lambda_J \leftarrow \mathrm{eigenvalues}(\mathbf{M}^{(\ell)}_J)$\;
     $d_N^{\ell} \leftarrow N! (2 \ell + 1) / [(N / 2 + \ell + 1)! (N / 2 - \ell)!]$\;
     $S_J \leftarrow S_J - \sum_{j = 0}^{2 \ell + 1} \lambda_J(j) \ln(\lambda_J(j) / d^{\ell}_N)$\;
}
\caption{The algebraic entanglement entropy algorithm for the spin degree of freedom $J$ of a mixed state $\hat{\rho}$.
Similar to Algorithm~\ref{alg:EE}, the starting state for layer $\ell$ is found by Gram-Schmidt orthogonalization using all previous states that have $m_j = m_k = \ell$.
For the momentum degree of freedom $K$, the for loop over $j,j',k$ becomes a for loop over $k,k',j$, the for loop over $j,j'$ becomes a for loop over $k,k'$, and the sums $\sum_{j = 0}^{2 \ell + 1}$ become $\sum_{k = 0}^{2 \ell + 1}$.}
\end{algorithm}

For mixed states, the entanglement entropy of just $\hat{\rho}_J$ or $\hat{\rho}_K$ is often not very illuminating as it could simply indicate entanglement between the degrees of freedom and an external environment rather than algebraic entanglement between the two degrees of freedom. 
However, our algorithm allows us to calculate any other quantities depending on the spectrum of $\hat{\rho}_J$ and $\hat{\rho}_K$.
In particular, we can calculate the coherent information~\cite{Wilde},
\begin{equation} \label{eq:coherentInformation}
\begin{aligned}
    I (J\rangle K) &= S_E [\hat{\rho}_K] - S_E [\hat{\rho}], \\
    I (K\rangle J) &= S_E [\hat{\rho}_J] - S_E [\hat{\rho}],
\end{aligned}
\end{equation}
which we note is the negative of the conditional entropy~\cite{Cerf,Horodecki2}.
Therefore, the coherent information provides a condition for state transfer whose conventional explanation (from Ref.~\cite{Horodecki2}) goes as follows.
Imagine there are two parties, $J$ and $K$. 
Party $J$ has access to the full $\hat{\rho}$ while party $K$ only has access to $\hat{\rho}_K$. 
A negative $I(J\rangle K)$ can be understood as the number of quantum bits that party $J$ must send to $K$ in order to achieve state transfer.
If it is positive, however, then the sender and receiver gain access to additional quantum resources after state transfer.
Put more simply, coherent information measures quantum correlations in a similar way as mutual information measures classical correlations~\cite{Wilde}.
Therefore, the positivity of~\cref{eq:coherentInformation} provides a sufficient condition for algebraic entanglement in mixed states~\cite{Cerf,Horodecki2}.
Interpreted in the context of motional and internal degrees of freedom, this means directly teleporting the internal state of the atoms onto their momentum degree of freedom is possible~\cite{Chen}, in principle.

\section{Examples} \label{sec:examples}
We will now calculate the algebraic entanglement entropy for four example physical systems.
First, we will calculate the algebraic entanglement entropy for the spin and momentum degrees of freedom for a state evolved under Eq.~\eqref{H_E} and compare the algorithm to the analytical solution from Sec.~\ref{sec:IllustrativeExample}.
Second, we will revisit the example of Refs.~\cite{Wilson,Christopher_Wilsons_PaperAnd_Definitely_Not_Jarrods_Or_Johns} and compare the algorithm to a simulation using the full $4^N$ dynamics.
Third, we will introduce decoherence into the model from Refs.~\cite{Christopher_Wilsons_PaperAnd_Definitely_Not_Jarrods_Or_Johns,Wilson}, where the loss is caused by a leaky cavity.
Lastly, we will calculate the relevant entropy dynamics for a model with dynamics driven by solely decoherence from our companion paper, Ref.~\cite{Reilly}.
The last three examples, Secs.~\ref{sec:BOAT}-~\ref{sec:SU4super}, all require a clever application of representation theory in order to simulate anything more than a few particles. 
In all four examples, we let the spin degree of freedom correspond to $J$ while momentum corresponds to $K$.

\subsection{Illustrative Example, Revisited.}\label{sec:example1}
\begin{figure}
    \centerline{\includegraphics[width=\linewidth]{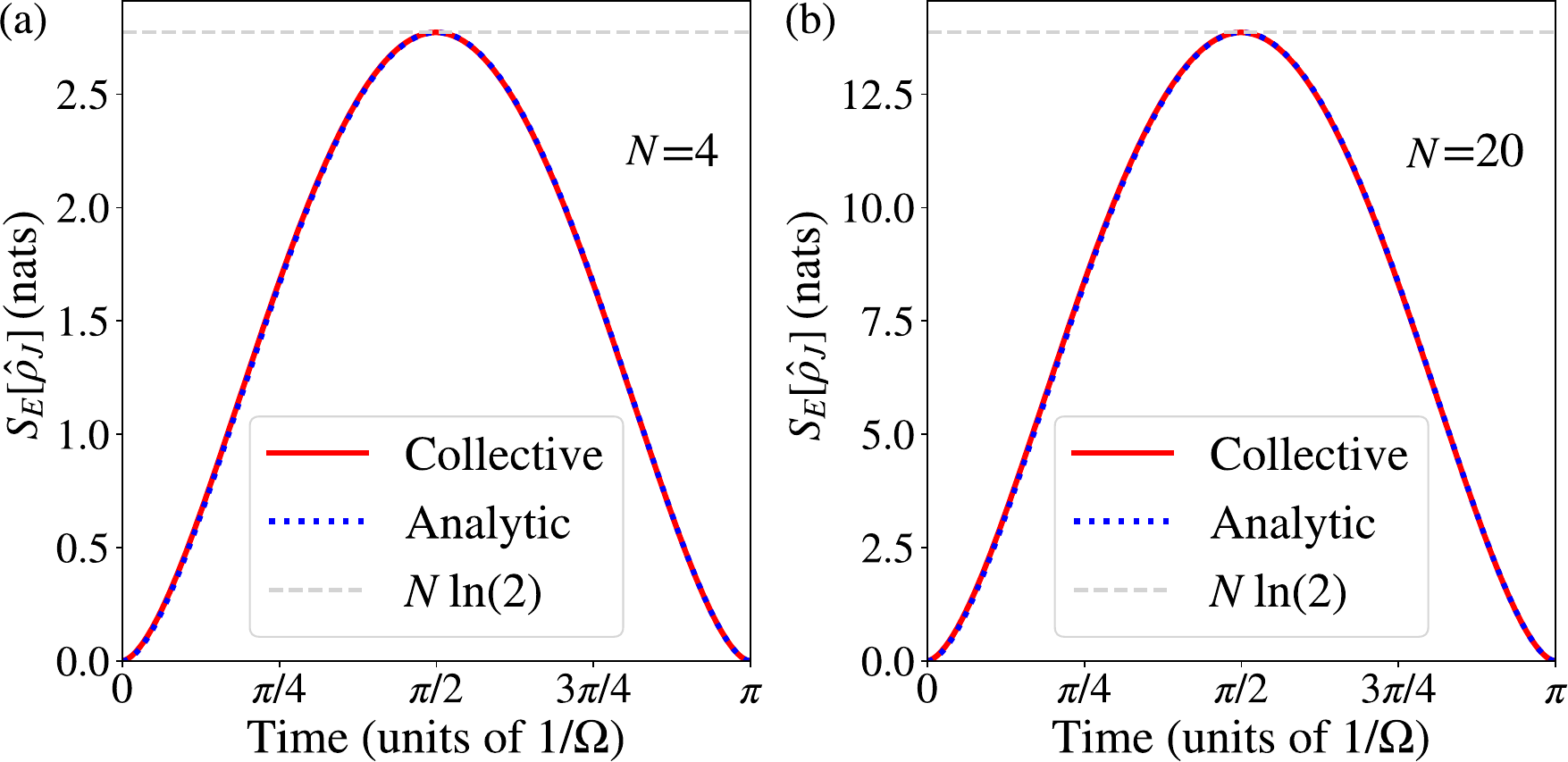}}
    \caption{
    The algebraic entanglement entropy for the collective internal states (corresponding to $J$) of the atoms for the Hamiltonian in Eq.~\eqref{eq:H_E2}. 
    (a) shows results for $N = 4$ while (b) shows results for $N = 20$. 
    We display the numerical results (red lines) and the analytical results from Eq.~\eqref{eq:AnalyticSolution} (blue lines).
    In both plots, we see that the entanglement entropy grows to the maximum possible value of $N \ln(2)$ (gray lines) given in Eq.~\eqref{eq:S_E_max} at $t = \pi / ( 2 \Omega)$.
    We find that the results from Algorithm~\ref{alg:EE} for the algebraic entanglement entropy follows the analytical solutions exactly.
    }
    \label{fig:Ex_plot}
\end{figure}

We will start with the illustrative example from Sec.~\ref{sec:IllustrativeExample}.
We see Eq.~\eqref{H_E} has the form of the operators given in Eq.~\eqref{su4Generators}, namely,
\begin{equation} \label{eq:H_E2}
    \hat{H}_{\mathrm{IE}} = \frac{\hbar \Delta}{2} \hat{G}_{z I} + \frac{\hbar \Omega}{2} \hat{G}_{x x}. 
\end{equation}
From Eq.~\eqref{eq:analyticSol}, we can find the algebraic entanglement entropy analytically.
For sake of simplicity, we will once again consider the limit of zero detuning, $\Delta = 0$, where the algebraic entanglement entropy is given by
\begin{equation} \label{eq:AnalyticSolution}
    \begin{aligned}
S_E [\hat{\rho}_J] = & - N \cos^2\left( \frac{\Omega t}{2} \right) \ln\left[ \cos^2\left( \frac{\Omega t}{2} \right) \right] \\ 
& - N \sin^2\left( \frac{\Omega t}{2} \right) \ln\left[ \sin^2\left( \frac{\Omega t}{2} \right) \right],
    \end{aligned}
\end{equation}
and $S_E [\hat{\rho}_K] = S_E [\hat{\rho}_J]$.
To compare with Algorithm~\ref{alg:EE}, we evolve the system from its initial state numerically and calculate the algebraic entanglement entropy using the steps outlined in Sec.~\ref{sec:entropy}.
The results are shown in Fig.~\ref{fig:Ex_plot}.

In Fig.~\ref{fig:Ex_plot}, we see that an algorithm that uses a Hilbert space that grows polynomially with $N$ can still accurately calculate an entanglement entropy which grows proportional to $N$ rather than with $\ln(N)$.
This is only possible because we have utilized the multiplicity $d^{\ell}_N$ of the irreps of $\SU{2}$.
This is interesting because it shows that these $\SU{4}$ dynamics are efficiently simulatable in a polynomial-scaled space, and the dynamics can generate states with algebraic entanglement entropy growing faster than the dimension of the Hilbert space.
In other words, the algebraic entanglement within the basis states allows algebraic entanglement entropy growth faster than $\ln[\dim(\mathcal{H}_\mathrm{sym})]$, which is what one would expect from a maximally mixed state whose eigenvalues are $1 / \dim(\mathcal{H}_{\mathrm{sym}})$.

\subsection{BOAT model} \label{sec:BOAT}
\begin{figure}
    \centerline{\includegraphics[width=\linewidth]{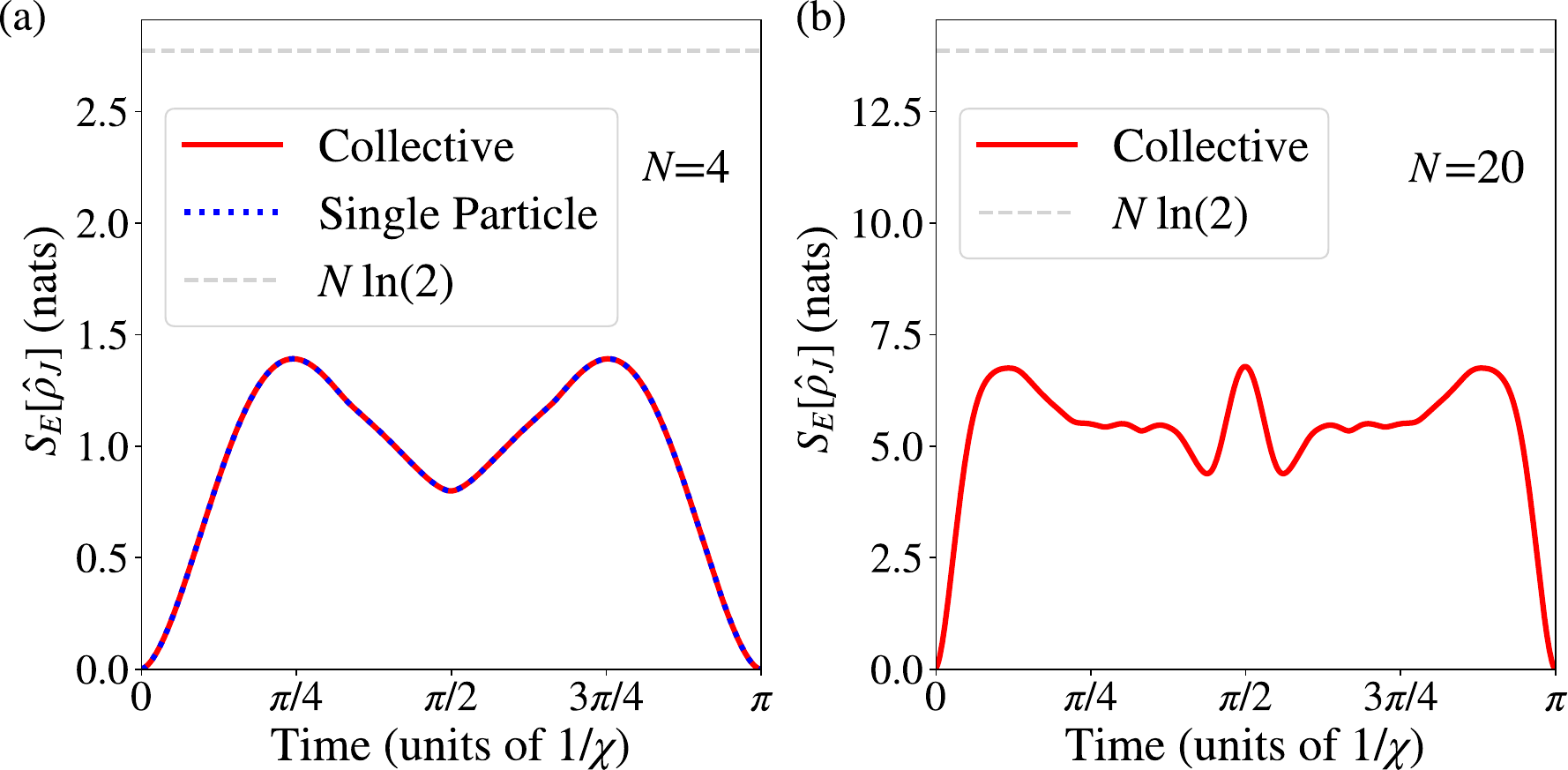}}
    \caption{
    The algebraic entanglement entropy for the collective internal states (corresponding to $J$) of the atoms for the BOAT Hamiltonian Eq.~\eqref{eq:H_BOAT}. 
    (a) shows results for $N = 4$. 
    Here, we can simulate the dynamics in the full single-particle basis and calculate the algebraic entanglement entropy (blue line), where we find it exactly matches the collective simulation (red line).
    (b) shows results for $N = 20$.
    Here, simulating the full $4^N$ dynamics is infeasible. 
    However, the algorithm which runs on the $\mathcal{O}(N^3)$ state space still calculates the algebraic entanglement entropy efficiently.}
    \label{fig:EpEm_plot}
\end{figure}

We now revisit the Hamiltonian presented in Refs.~\cite{Wilson,Christopher_Wilsons_PaperAnd_Definitely_Not_Jarrods_Or_Johns}, dubbed the beyond one-axis twisting (BOAT) model.
Here, both interparticle entanglement and algebraic entanglement are generated, which makes analytical calculations used in the previous example no longer tractable. 
The Hamiltonian is given by
\begin{equation} \label{eq:H_BOAT}
    \hat{H}_{\mathrm{BOAT}} = \hbar \chi \hat{G}_{(+) x} \hat{G}_{(-) x},
\end{equation}
which corresponds to a one-axis-twisting Hamiltonian~\cite{Kitagawa} with the additional complication that the atoms have two momentum states. 
This Hamiltonian can be accomplished by a particle beam traversing a dispersive cavity while energy requirements limits the particles momenta along the cavity axis to $\pm \hbar k / 2$~\cite{Wilson}. 
After adiabatically eliminating the cavity field, we arrive at the atom-only nonlinear Hamiltonian in Eq.~\eqref{eq:H_BOAT}.
The nonlinearity in this model will generate interparticle entanglement while the structure of the operators will generate algebraic entanglement, the combination of which allows for multi-parameter estimation beyond the standard quantum limit~\cite{Christopher_Wilsons_PaperAnd_Definitely_Not_Jarrods_Or_Johns}.
Since the Hamiltonian is described fully by collective $\SU{4}$ operators, the dynamics can be characterized by the states presented in Sec.~\ref{sec:IllustrativeExample}.
We assume the atoms are initially in a superposition of the ground and excited state with a momenta of $- \hbar k/2$, 
\begin{equation} \label{eq:psi_start}
    \ket{\Psi (0)} = \bigotimes_{j = 1}^N \left( \left[ \frac{ \ket{0}_j + \ket{1}_j }{\sqrt{2}} \right] \otimes \down_j \right).
\end{equation}
We can evolve the system using the Schr\"odinger equation and calculate the algebraic entanglement entropy $S_E [\hat{\rho}_J] = S_E [\hat{\rho}_K]$.
The numerical results for the dynamics of the algebraic entanglement entropy for $N = 4$ are shown in Fig.~\ref{fig:EpEm_plot}(a).
In order to benchmark the algorithm, we also simulate dynamics in the full $4^N$ single-particle basis.
We find perfect agreement between the two results. 

We take a moment to highlight an interesting fact.
If we were to calculate an entanglement witness such as the quantum Fisher information for $\hat{G}_{x I}$, as is done in Ref.~\cite{Wilson}, we would find that there is a large amount of interparticle entanglement.
Despite this, we see that there is less algebraic entanglement than in the previous example shown in Sec.~\ref{sec:example1}.
In fact, we can see from the structure of the Hamiltonian in Eq.~\eqref{eq:H_E2} and its solutions that there is no interparticle entanglement whatsoever in the previous example, despite there being a large amount of algebraic entanglement, which is how we are able to easily find analytical solutions to the previous example.
This is no longer the case for the BOAT model, and so the results of Fig.~\ref{fig:EpEm_plot}(b) for $N = 20$ would not be possible without the polynomial-scaling of Algorithm~\ref{alg:EE}. 
Therefore, we are able to efficiently perform a calculation in Fig.~\ref{fig:EpEm_plot}(b) that would typically require a $4^{20} \sim 10^{12}$ dimensional Hilbert space.  

\subsection{Leaky BOAT Model} \label{sec:LeakyBOAT}
\begin{figure*}
    \centerline{\includegraphics[width=0.8\linewidth]{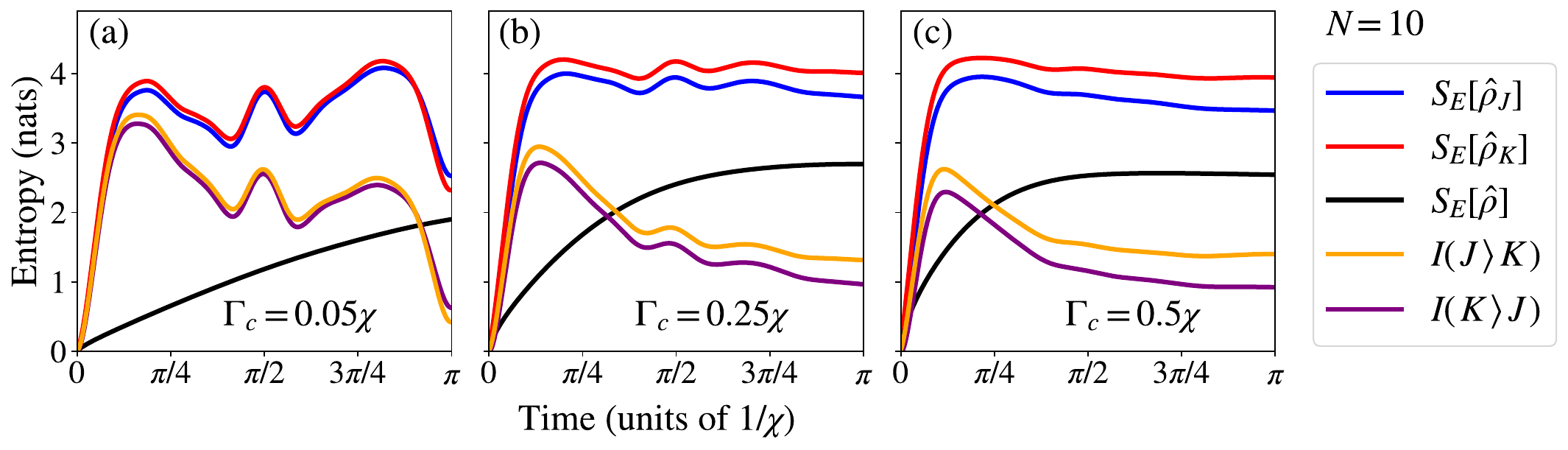}}
    \caption{ 
    Results for the leaky BOAT model Eq.~\eqref{eq:MasterEq}.
    We display the algebraic entanglement entropy for $\hat{\rho}_J$ (blue) and $\hat{\rho}_K$ (red), the entropy for the total state $\hat{\rho}$ (black), and the coherent informations for $I(J \rangle K)$ (orange) and $I(K \rangle J)$ (purple). 
    The parameters are (a) $\Gamma_c = 0.05 \chi$, (b) $\Gamma_c = 0.25 \chi$, and (c) $\Gamma_c = 0.5 \chi$.
    In all three, we see that both $I(J\rangle K)$ and $I(K \rangle J)$ are positive, indicating the presence of algebraic entanglement between $J$ and $K$.
    For larger decay rates, we see that the entanglement entropy of $\hat{\rho}_J$ and $\hat{\rho}_K$ differ due to entanglement with the environment, which is indicated by the growing entropy of $\hat{\rho}$.}
    \label{fig:LeakyBOAT_plot}
\end{figure*}
We now introduce an additional complication not included in the BOAT model in Refs.~\cite{Wilson,Christopher_Wilsons_PaperAnd_Definitely_Not_Jarrods_Or_Johns}, which we use to highlight the need to calculate coherent information in the case of open quantum systems.
If the decay of the cavity field into free space were included in the model of Ref.~\cite{Wilson}, one would find this corresponds to a decoherence jump operator that goes as $\hat{L}_d = \sqrt{\Gamma_c} \, \hat{G}_{(-) x}$ after adiabatically eliminating the cavity field, similar to Ref.~\cite{Reilly}.
Here, time dynamics are dictated by the Born-Markov master equation,
\begin{equation} \label{eq:MasterEq}
    \pd{\hat{\rho}} = \hat{\mathcal{L}}_{\mathrm{BOAT}} \hat{\rho} = - \frac{i}{\hbar} \left[ \hat{H}_{\mathrm{BOAT}}, \hat{\rho} \right] + \hat{\mathcal{D}} [\hat{L}_d] \hat{\rho},
\end{equation}
where $\hat{\mathcal{L}}_{\mathrm{BOAT}}$ is the Liouvillian superoperator, $\hat{H}_{\mathrm{BOAT}}$ is the Hamiltonian from Eq.~\eqref{eq:H_BOAT}, and the Lindblad superoperator is given by 
\begin{equation}
    \hat{\mathcal{D}} [ \hat{O} ] \hat{\rho} = \hat{O} \hat{\rho} \hat{O}^{\dagger} - ( \hat{O}^{\dagger} \hat{O} \hat{\rho} + \hat{\rho} \hat{O}^{\dagger} \hat{O} ) / 2.
\end{equation}
We simulate the dynamics for this model using the same starting state given by~\cref{eq:psi_start} and three values of decoherence, $\Gamma_c = 0.05 \chi$, $\Gamma_c = 0.25 \chi$, and $\Gamma_c = 0.5 \chi$, shown in Figs.~\ref{fig:LeakyBOAT_plot}(a), (b), and (c), respectively.
For these dynamics, we also use~\cref{eq:coherentInformation} to calculate the coherent information between the two degrees of freedom to demonstrate algebraic entanglement.
As expected, we find that the general features of the closed system in Fig.~\ref{fig:EpEm_plot}, namely the three ``humps'' where the entanglement entropy spikes to its maximum value in Fig.~\ref{fig:EpEm_plot}(b), decay away as the decay rate $\Gamma_c$ increases compared to the squeezing rate $\chi$. 
Moreover, Fig.~\ref{fig:LeakyBOAT_plot}(a) demonstrates the importance of studying the coherent information rather than just the entanglement entropy when studying open quantum systems. 
We see that maximum value of the three ``humps'' in the entanglement entropy increases as time increases, but ``humps'' in the coherent information decreases as time increases.
This indicates that the algebraic entanglement between the two degrees of freedom is decreasing as the entanglement of the total system with the environment is increasing over the dynamics. 

\begin{figure}
    \centerline{\includegraphics[width=\linewidth]{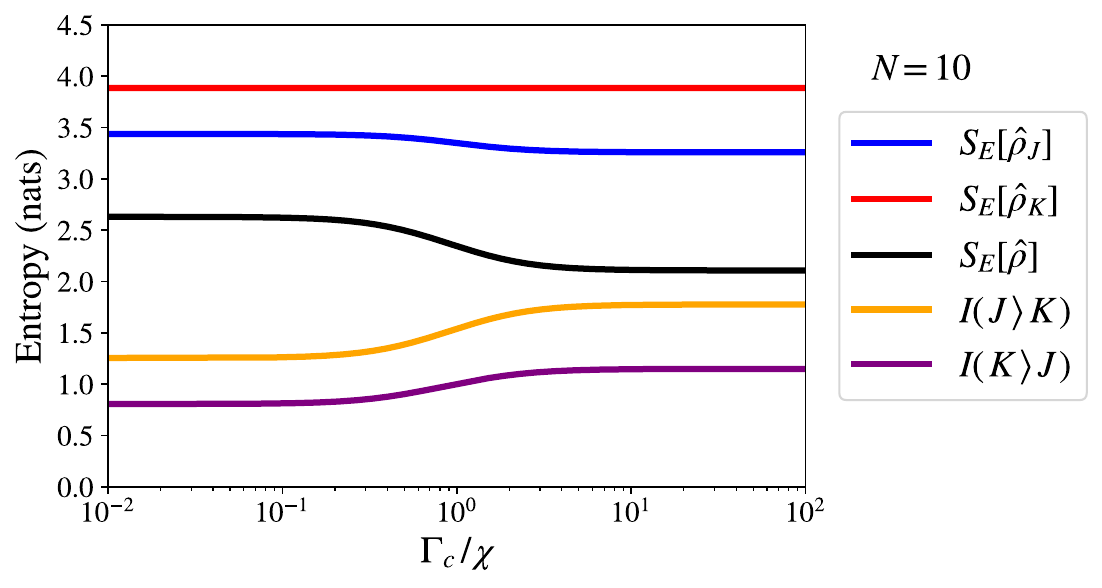}}
    \caption{
    The steady-state of the leaky BOAT model Eq.~\eqref{eq:MasterEq} as a function of $\Gamma_c / \chi$.
    We again plot the algebraic entanglement entropy for $\hat{\rho}_J$ (blue) and $\hat{\rho}_K$ (red), the entropy for the entire $\hat{\rho}$ (black), and the coherent informations $I(J \rangle K)$ (orange) and $I(K \rangle J)$ (purple).
    We see that, even when photon loss dominates dynamics ($\Gamma_c \gg \chi$), we still find positive coherent information indicating usable algebraic entanglement.}
    \label{fig:LeakyBOAT_SteadyState_plot}
\end{figure}

Lastly, we examine the entropies and coherent information of the model at steady-state, $\hat{\mathcal{L}}_{\mathrm{BOAT}} \hat{\rho}_{\mathrm{ss}} = 0$, as a function of $\chi$ and $\Gamma_c$.
To find the state which~\cref{eq:psi_start} relaxes into, we simply simulate the dynamics for a long time $\hat{\rho}_{\mathrm{ss}} = \hat{\rho}(t\to\infty)$.
The results are shown in~\cref{fig:LeakyBOAT_SteadyState_plot}.
We see that the steady-state entropy on the $K$ degree of freedom is fixed regardless of $\Gamma_c$ and $\chi$ even though it varies during the dynamics [Fig.~\ref{fig:LeakyBOAT_plot}].
On the $J$ degree of freedom, however, we see that steady-state entropy decreases as the ratio $\Gamma_c/\chi$ goes from below one to above one.
The fixed entropy of the $K$ degree of freedom in steady-state is a result of the fact that both $\hat{H}_{\mathrm{BOAT}}$ and $\hat{L}_d$ commute with the operator $\hat{E}^2 \equiv \hat{G}_{(+) x} \hat{G}_{(-) x} + \hat{G}_{z I}^2 - \hat{G}_{z I}$, meaning a strong symmetry is present in the Liouvillian~\cite{Buca,Lieu,Nairn}.
This symmetry corresponds to the Casimir operator for the su(2) sub-algebra $\{ \hat{G}_{z I},\hat{G}_{(+) x},\hat{G}_{(-) x} \}$, which fixes the steady state dynamics for $K$ based on the starting state.
The entropy of the $J$ degree of freedom in steady state, meanwhile, changes because the Hamiltonian effectively ``re-pumps'' the atoms, thereby allowing more entanglement to be generated with the environment via emission.
This means each atom is near the equator of the $J$-Bloch sphere and, in terms of~\cref{fig:Pyramid} (b), the density matrix will have support across multiple rows of $m_j$.

\subsection{Spin-Momentum Superradiance} \label{sec:SU4super}
As a final example, we consider the dual superradiance system from Ref.~\cite{Reilly} as an example of a fully dissipative system that generates algebraic entanglement between $J$ and $K$.
This model begins with the leaky BOAT Liouvillian Eq.~\eqref{eq:MasterEq} that is on resonance, such that $\chi = 0$, mediated by a bad-cavity along the $x$-axis.
A second bad-cavity along the $z$-axis and a classical drive can mediate a two-photon Raman transition between the two internal states mediated by an auxiliary excited state~\cite{Reilly,Reilly2}. 
This leads to collective pumping from ground to excited state at a rate $W$
without changing the momentum in the $x$-direction, such that the jump operator is $\hat{L}_r = \sqrt{W} \hat{G}_{(+) I}$.
The master equation is thus
\begin{equation} \label{eq:SuperradMasterEq}
    \pd{\hat{\rho}} = \hat{\mathcal{L}}_{\mathrm{SMS}} \hat{\rho} = \hat{\mathcal{D}} [\hat{L}_d] \hat{\rho} + \hat{\mathcal{D}} [\hat{L}_r] \hat{\rho},
\end{equation}
which, notably, has no unitary dynamics.
We once again consider the atoms starting in the state given by~\cref{eq:psi_start}. 
However, it is important to note that this initial state leads to the same steady state as the one considered in Ref.~\cite{Reilly}, $\ket{\Psi' (0)} = \bigotimes_j^N (\ket{0}_j \otimes \ket{\downarrow}_j)$, because both initial states are in the same $\mathrm{U} (1)$ subgroup, although the dynamics may be slightly different.

\begin{figure}
    \centerline{\includegraphics[width=\linewidth]{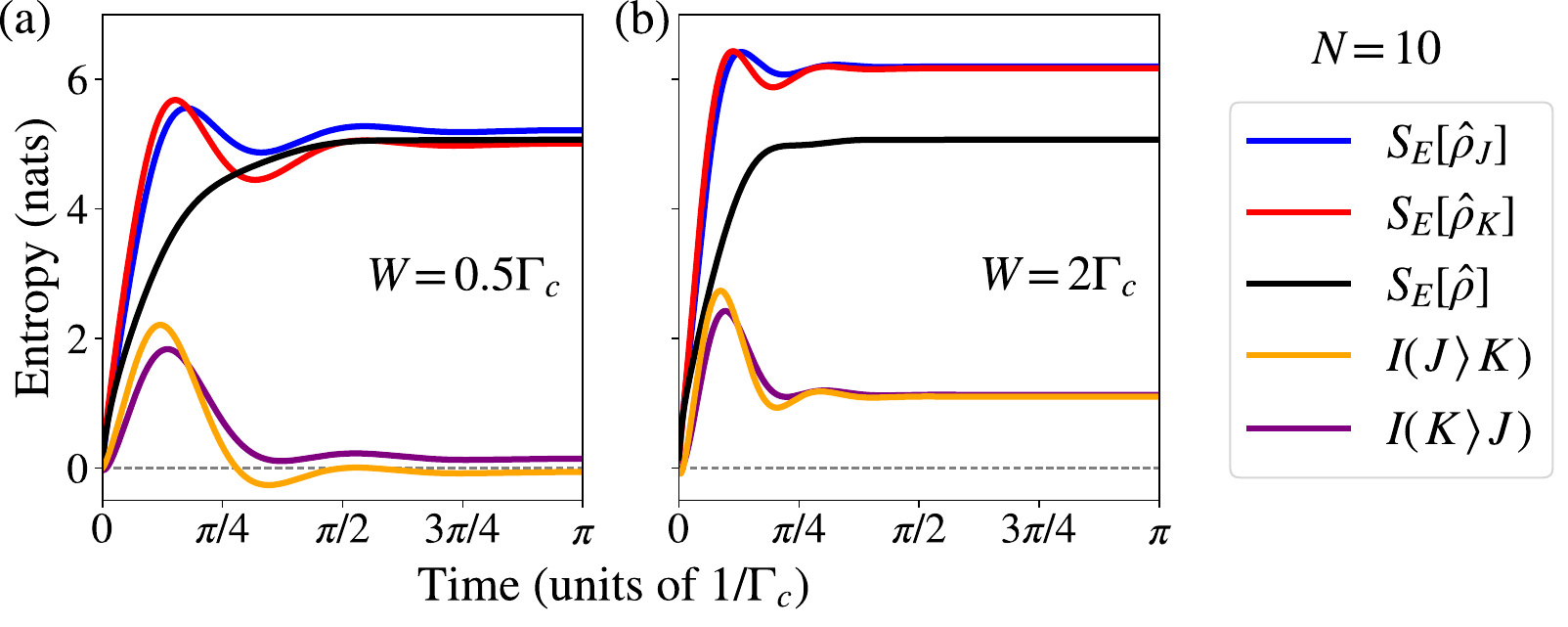}}
    \caption{
    The dynamics of the spin-momentum superradiance system from Eq.~\eqref{eq:SuperradMasterEq}. 
    We display the algebraic entanglement entropy for $\hat{\rho}_J$ (blue) and $\hat{\rho}_K$ (red), the entropy for the entire $\hat{\rho}$ (black), and the coherent informations $I(J \rangle K)$ (orange) and $I(K \rangle J)$ (purple). 
    The parameters are (a) $W = 0.5 \Gamma_c$ and (b) $W = 2 \Gamma_c$.}
    \label{fig:SMS_plot}
\end{figure}

We simulate the dynamics of the various entropy quantities for $W = \Gamma_c / 2$ and $W = 2 \Gamma_c$, which are shown in~\cref{fig:SMS_plot} (a) and (b), respectively.
We see that the algebraic entanglement entropy for $J$ and $K$ each initially grow faster than the total entropy, meaning we initially find a positive coherent information for both degrees of freedom and values of $\Gamma_c$.
This suggests that dissipative engineering of these systems with $\SU{4}$ symmetry provide powerful quantum information processing platforms.
Surprisingly, we find that the case of a larger incoherent pump rate, $W = 2 \Gamma_c$, generates more coherent information at steady-state than that of a lower incoherent pump rate, $W = \Gamma_c / 2$. 
This is because the higher pump rate allows the system to relax into a state with population spread across the multiple $\ell$ layers of the pyramid.

\begin{figure}
    \centerline{\includegraphics[width=\linewidth]{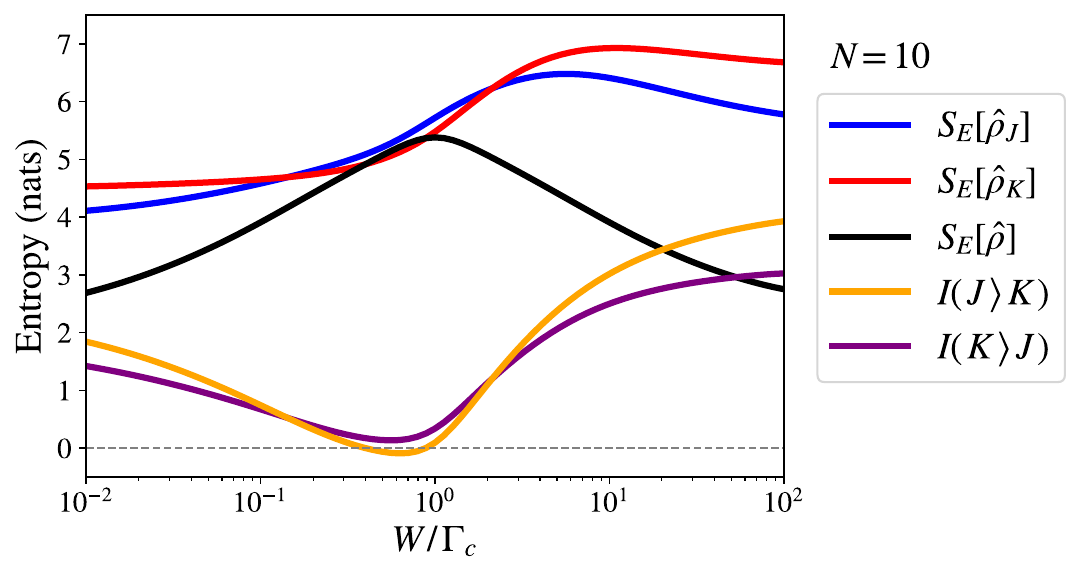}}
    \caption{ 
    The steady-state entropies of the spin-momentum superradiance system of Eq.~\eqref{eq:SuperradMasterEq} as a function of $W / \Gamma_c$.
    We see that fully dissipative dynamics still create large positive coherent information, and thus algebraic entanglement between $J$ and $K$.}
    \label{fig:CrossedCav_SteadyState_plot}
\end{figure}

We now examine the entropy quantities at steady-state, $\hat{\mathcal{L}}_{\mathrm{SMS}} \hat{\rho}_{\mathrm{ss}} = 0$, as a function of $W$ compared to $\Gamma_c$.
To find the state which~\cref{eq:psi_start} relaxes into, we again simulate the dynamics for a long time $\hat{\rho}_{\mathrm{ss}} = \hat{\rho}(t\to\infty)$.
The results are shown in~\cref{fig:CrossedCav_SteadyState_plot}.
There, we see a more complicated response of the coherent information to the ratios of the re-pump and decay rates.
The operator $\hat{E}^2 = \hat{G}_{(+) x} \hat{G}_{(-) x} + \hat{G}_{z I}^2 - \hat{G}_{z I}$ does not commute with $\hat{G}_{(+) I}$, and therefore we see the entropy on $K$ differ as we adjust the rates.
For the value $W=\Gamma_c$, we see that the system reaches its maximum entanglement with the environment, with both the entropy of the $J$ and $K$ degrees of roughly that of the total entropy.
For both $W>\Gamma_c$, and $W<\Gamma_c$, we see that the coherent information grows, while the total entropy of the system decreases.
Using the steady-state results presented in Ref.~\cite{Reilly}, we can gain insight into the algebraic entanglement generated in the system. 
Namely, in the case $W \gg \Gamma_c$, one finds that the steady-state dipole length of $J$ becomes minuscule, $\exv{\hat{J}^2}_{\mathrm{ss}} \ll (N / 2) (N / 2 + 1)$ and so $j \ll N / 2$.
Since $\ell = j = k$, this indicates that there are $(N / 2 - \ell) \gg 1$ many singlet pairs in both the $J$ and $K$ degrees of freedom, and so the basis states of $\ket{\ell,m_j,m_k}$ that are populated are endowed with a massive amount of algebraic entanglement, and this explains why the regime $W \gg \Gamma_c$ has the maximal coherent information between the two degrees of freedom. 

\section{Conclusion} \label{sec:Conclusion}
In this paper, we explored the connection between algebraic entanglement between degrees of freedom and Lie group symmetries. 
We have shown that one can exploit the structure of the irreps of compact Lie groups to perform operations in polynomial scaling that would typically require an exponentially-scaled complexity.
In particular, we exploit the structure of a direct sum between irreps of $\SU{2}$ to diagonalize density matrices in a block-by-block manner, the largest block scales as $\mathcal{O}(N^2)$, and then exploit the multiplicity of the irreps to achieve a linear scaling of entropy with a Hilbert space whose dimensionality only has polynomial scaling. 
While we have focused on the case $\SU{2} \otimes \SU{2} < \SU{4}$ here, our methodology can be readily generalized to $\SU{n} \otimes \SU{n} < \SU{n^2}$ systems using multiplet diagrams and Young tableaux to systematically explore irreps, as well as the simulation basis from Ref.~\cite{Bastin} for the actions of collective operators. 
Future work will be dedicated to extending the ideas from our work to $\SU{m} \otimes \SU{n} < \SU{m \cross n}$ systems, and then $\SU{\ell} \otimes \SU{m} \otimes \SU{n} \otimes \ldots$ systems by tracing out subsystems one-by-one with a suitable adaptation of the pyramid that explores all the irreps of the underlying degrees of freedom (e.g., utilizing Ref.~\cite{Bastin}).
This could directly relate to the exact calculation of algebraic entanglement entropy between different ensembles of superspins in partially permutational symmetric systems using the simulation technique of Ref.~\cite{Lee}.

The most immediate application of our procedure is to quantum information science. 
Here, entanglement entropy is an indication that certain information processes can be performed, such as quantum teleportation~\cite{Chen} or device independent quantum key distribution for cryptography~\cite{Pirandola}.
This can offer insight into the role and advantages of algebraic entanglement between degrees of freedom in quantum computing algorithms. 
Calculation of algebraic entanglement entropy may shed insight into entanglement swapping between intraparticle and interparticle entanglement~\cite{Tang,Adhikari,Bayal}, and our work may inspire more general swapping protocols, e.g., swapping general algebraic entanglement into interparticle entanglement in a single degree of freedom~\cite{Takeda,Yang}.
Moreover, our procedure could help explore the role of quantum entanglement in well-known processes where entanglement is not usually thought of.
For example, the role of spontaneous emission, i.e., entanglement with free-space electromagnetic modes, in laser cooling has recently been scrutinized~\cite{Metcalf,Corder,Bartolotta,Reilly4}, and generalizations of our procedure could shed insight into the entanglement between the subsystems (particles and cooling light field) and its relation to phase space compression of the particles.
Lastly, there may be non-trivial connections between the rapid growth of algebraic entanglement entropy and information scrambling~\cite{Landsman2019,Wang}.
There, information is non-locally dispersed across subsystems and has important implications for quantum chaos and black holes~\cite{Hayden2007}.
Therefore, efficient determination of algebraic entanglement entropy provides another avenue to understanding how entanglement and correlations are distributed in complex quantum many-body systems.

\section*{Acknowledgements}
We would like to thank Simon B. J\"ager, Peter Zoller, John Cooper, Gage W. Harmon, Haonan Liu, Charles Marrder, and Lyryl H. C. Vaecairn for useful discussions.
This material is based upon work supported by the U.S. Department of Energy, Office of Science, National Quantum Information Science Research Centers, Quantum Systems Accelerator (Award No. DE-SCL0000121), and by the National Science Foundation Grant Nos.\ 2016244 and 2317149.

\appendix
\renewcommand{\thefigure}{\thesection\arabic{figure}}
\counterwithin{figure}{section}

\section{Classes of Entanglement} \label{sec:EntClasses}
\begin{figure}[t]
\centerline{\includegraphics[width=\columnwidth]{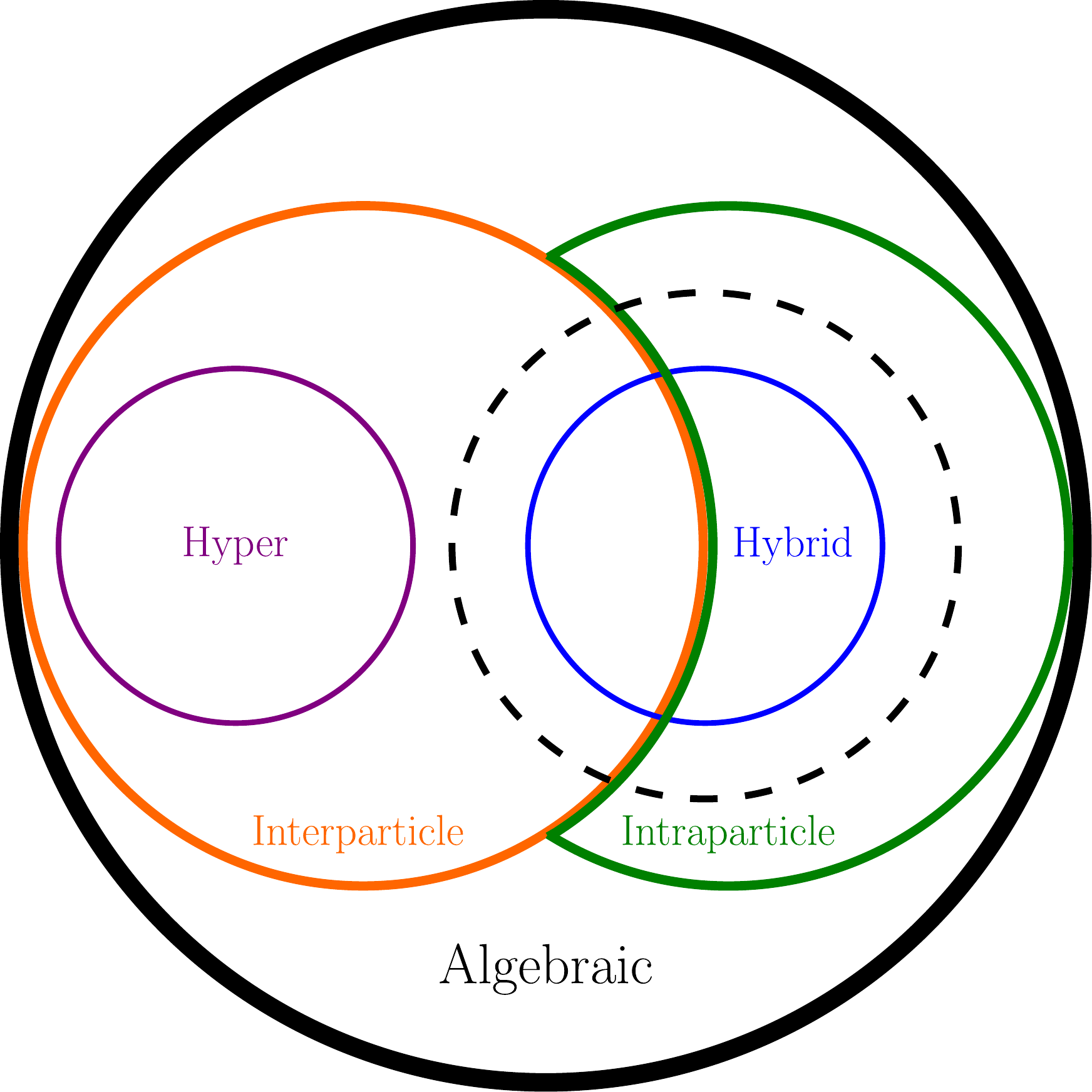}}
    \caption{A Venn diagram of the various classes of entanglement in quantum mechanics mentioned in Secs.~\ref{sec:Intro} and~\ref{sec:IllustrativeExample}.
    The dashed black line represents algebraic entanglement between distinct degrees of freedom, which is the class of entanglement that the algorithm in Sec.~\ref{sec:entropy} calculates. 
    Note that the Venn diagram refers to entanglement classifications and not sets of physical states (for which they may \emph{all} overlap for certain states).}
    \label{fig:VennDiag}
\end{figure}

In this appendix, we provide a more in-depth analysis of the various classes of entanglement discussed in the main text.
We emphasize that the algorithm in Sec.~\ref{sec:entropy} is tailored to calculate \emph{algebraic entanglement between degrees of freedom}.
A short summary of each class of entanglement with the relevant bi- or multipartition is given below (here, we use ``particle'' as a catch-all term for atoms, molecules, ions, etc.):
\begin{itemize}
    \item \textbf{Algebraic:} 
    In the context of typical quantum mechanical systems where the relevant von Neumann algebras are type I factors~\cite{vonNeumann2}, algebraic entanglement is entanglement with respect to a bi- or multipartition of the observable algebra into commuting subalgebras associated with distinct subsystems (e.g., degrees of freedom or parties).
    It therefore is an umbrella class that encompasses all the following ``particle-based'' entanglement classes, as well as particle-subsystem entanglement (e.g., particle-cavity entanglement),  entanglement between different collections of particles, and more.

    \item \textbf{Algebraic between degrees of freedom:}
    Entanglement with respect to a bi- or multipartition between distinct subalgebras associated with the observables of different degrees of freedom.
    It could therefore refer to inter- or intraparticle entanglement, as well as hybrid entanglement.
    This is the class of entanglement that the main text focuses on and is calculated with the algorithms in Sec.~\ref{sec:entropy}.

    \item \textbf{Interparticle:} 
    Entanglement with respect to a bi- or multipartition between the different particles in the quantum system.

    \item \textbf{Intraparticle:} 
    Entanglement that is typically defined with respect to a bi- or multipartition between the different degrees of freedom \emph{within} a given particle of the quantum system.
    In this sense, intraparticle entanglement is algebraic entanglement between distinct degrees of freedom.
    However, some systems do not admit a natural tensor product decomposition into degrees of freedom, e.g., entanglement between the valence electrons in an alkaline-earth-like atom when treated in its physical fermionic Hilbert space, and so the more general form of algebraic entanglement is required~\cite{Zanardi,Benatti}.

    \item \textbf{Hybrid:} 
    Entanglement with respect to a bi- or multipartition between degrees of freedom of a collection of particles in which at least one degree of freedom is discrete-variable while at least one degree of freedom is continuous-variable.
    It can therefore refer to either inter- or intraparticle entanglement.

    \item \textbf{Hyper:} 
    Entanglement with respect to a bi- or multipartition in which a collection of particles simultaneously has interparticle entanglement in more than one degree of freedom.
    Therefore, hyperentanglement is distinct from algebraic entanglement between degrees of freedom and so is not encapsulated by the algorithm in Sec.~\ref{sec:entropy}.
\end{itemize}

Figure~\ref{fig:VennDiag} shows a Venn diagram of the various classes of entanglement in which the dashed black line represents the algebraic entanglement between distinct degrees of freedom discussed in the main text.
While a state may have both hyperentanglement and hybrid entanglement (see, e.g., Ref.~\cite{Li2}), just as it may have both inter- and intraparticle entanglement, note that they are not overlapping in Fig.~\ref{fig:VennDiag} because they characterize different structural aspects of bi- or multipartite quantum correlations; an entanglement witness tailored to one class is not necessarily sensitive to the other.
We also note that algebraic entanglement, especially in the context of quantum field theory where one cannot in general rely on a simple tensor-product decomposition of the Hilbert space into subsystems, is a more general concept that extends naturally to general von Neumann algebras~\cite{vonNeumann,vonNeumann2} (W*-algebras~\cite{Schwartz}), including those with type II and type III factors.
Finally, we note that some literature use the terminology ``hybrid entanglement'' even when the distinct degrees of freedom are both discrete-variable, e.g., Refs.~\cite{Karimi,Chen2}; we adopt the more standard terminology (entanglement between discrete-variable and continuous-variable) from, e.g., Refs.~\cite{Li2,Andersen,Guccione}.

\section{The Single-Particle Operators} \label{sec:singleParticleOps}
We consider the single-particle operators (for particle $j$)
\begin{equation}
    \begin{aligned}
& \hat{\sigma}_j^{(+)} = \op{1}{0}_j, \quad \hat{\sigma}_j^{(-)} = \op{0}{1}_j, \quad \hat{\sigma}^x_j = \hat{\sigma}_j^{(+)} + \hat{\sigma}_j^{(-)}, \\
& \hat{\sigma}^y_j = i \left( \hat{\sigma}_j^{(-)} - \hat{\sigma}_j^{(+)} \right), \quad \hat{\sigma}^z_j = \left[ \hat{\sigma}_j^{(+)}, \hat{\sigma}_j^{(-)} \right], 
    \end{aligned}
\end{equation}
on the $J$ degree of freedom and
\begin{equation}
    \begin{aligned}
& \hat{s}_j^{(+)} = \op{\upp}{\ddown}_j, \quad \hat{s}_j^{(-)} = \op{\ddown}{\upp}_j, \quad \hat{s}^x_j = \hat{s}_j^{(+)} + \hat{s}_j^{(-)}, \\
& \hat{s}^y_j = i \left( \hat{s}_j^{(-)} - \hat{s}_j^{(+)} \right), \quad \hat{s}^z_j = \left[ \hat{s}_j^{(+)}, \hat{s}_j^{(-)} \right],
    \end{aligned}
\end{equation}
on the $K$ degree of freedom. 
Sums over these operators for each atom generates the operators in the respective $\mathfrak{su} (2)$ sub-algebras.

\section{Computational Complexity}
In this appendix, we analyze the computational complexity of the various steps of Algorithms~\ref{alg:EE} and~\ref{alg:EEmixed}.
The relevant scalings for the pure state algorithm are shown in the following table:
\begin{center}
\begin{tabular}{|c|c|}
\hline
Step & Scaling \\
\hline
\hline
Irrep decomposition & $\mathcal{O}(N^3)$ \\
Basis state construction & $\mathcal{O}(N^5)$ \\
Gram-Schmidt orthogonalization & $\mathcal{O}(N^3)$ \\
Evaluation of $p_\ell(j,k)$ & $\mathcal{O}(N^4)$ \\
Eigenvalue decompositions & $\mathcal{O}(N^4)$ \\
\hline
Dominant term & $\mathcal{O}(N^5)$ \\
\hline
\end{tabular}
\end{center}
The relevant scalings for the mixed state algorithm are shown in the following table:
\begin{center}
\begin{tabular}{|c|c|}
\hline
Step & Scaling \\
\hline
\hline
Irrep decomposition & $\mathcal{O}(N^3)$ \\
Basis state construction & $\mathcal{O}(N^5)$ \\
Gram-Schmidt orthogonalization & $\mathcal{O}(N^3)$ \\
Evaluation of $P(j,j',k)$ & $\mathcal{O}(N^6)$ \\
Reduced matrix formation & $\mathcal{O}(N^3)$ \\
Eigenvalue decompositions & $\mathcal{O}(N^4)$ \\
\hline
Dominant term & $\mathcal{O}(N^6)$ \\
\hline
\end{tabular}
\end{center}

\bibliography{references.bib}

\end{document}